\documentclass[aps,prb,reprint,amsmath,amssymb,10pt,raggedbottom,superscriptaddress,floatfix]{revtex4-2}
\usepackage{graphicx}
\usepackage{bm}
\usepackage{placeins}
\usepackage[
  colorlinks=true,
  linkcolor=blue,
  citecolor=blue,
  urlcolor=blue,
  breaklinks=true
]{hyperref}

\newcommand{\AM}{\mathrm{AM}}
\newcommand{\BdG}{\mathrm{BdG}}


\begin{document}
\title{Majorana corner modes and tunable braiding in an altermagnet–superconductor heterostructure}
\author{Yingli Gao}
\affiliation{School of Science, Xi'an Technological University, Xi'an 710032, China}
\author{Kai Zhang}
\affiliation{School of Science, Xi'an Technological University, Xi'an 710032, China}
\author{Ling-Zheng Meng}
\affiliation{School of Science, Xi'an Technological University, Xi'an 710032, China}
\author{Junpeng Hou}
\affiliation{Pinterest Inc., San Francisco, California 94103, USA}
\author{Ning Li}
\affiliation{School of Science, Xi'an Technological University, Xi'an 710032, China}
\author{Ya-Jie Wu}
\email{wuyajie@xatu.edu.cn}
\begin{abstract}
Zeeman fields are usually required for realizing and manipulating Majorana corner modes at the cost of suppressed superconducting pairing. Altermagnetism offers a new route toward realizing Majorana corner modes under zero net external magnetic fields. Here we propose a theoretical scheme to host Majorana corner modes in a honeycomb-lattice altermagnet–superconductor heterostructure. In contrast to previous studies on square lattices, altermagnetism yields distinct effective low-energy edge masses on various inequivalent edges of the honeycomb lattice. Within a low-energy edge-theory framework, we establish the conditions under which Majorana corner modes emerge for both bearded and zigzag edge terminations. We further explicitly construct a complete set of Clifford gates by tuning the altermagnetic anisotropy and chemical potential of the heterostructure. Our results provide a promising platform for the realization and manipulation of Majorana corner modes. 
\end{abstract}
\maketitle
\makeatletter
\raggedbottom
\let\manuscript@firstpagebottom\@textbottom
\flushbottom
\let\manuscript@laterpagebottom\@textbottom
\def\@textbottom{%
  \ifnum\c@page=\@ne
    \manuscript@firstpagebottom
  \else
    \manuscript@laterpagebottom
  \fi
}
\makeatother

\makeatletter
\def\raggedcolumn@sw{%
  \ifnum\c@page=\@ne
    \expandafter\@firstoftwo
  \else
    \expandafter\@secondoftwo
  \fi
}
\makeatother

\section{Introduction and motivation}
Majorana zero modes (MZMs) serve as building blocks for topological qubits, and their coherent control enables the implementation of topological quantum gates \cite{ref1,ref2}. The pursuit and manipulation of MZMs therefore constitute a central objective in topological quantum computation. Early theoretical proposals for MZM realization were rooted in two-dimensional $p$-wave superconductors and the one-dimensional Kitaev chain \cite{ref3,ref4}. Nevertheless, intrinsic $p$-wave pairing remains challenging to achieve in both condensed-matter systems and ultracold atomic gases. To engineer effective $p$-wave pairing, numerous theoretical schemes have been put forward leveraging the synergy between spin-orbit coupling and $s$-wave superconductor pairing with Zeeman fields \cite{ref5,ref6,ref7,ref8}. Substantial progress has been made in the theoretical description and control of Majorana qubits \cite{ref9,ref10,ref11,ref12,ref13,ref14}. These topological-superconductor platforms generally require Zeeman fields to break time-reversal symmetry, and braiding operations for MZMs in two-dimensional systems typically rely on manipulating magnetically induced vortices.

Higher-order topological superconductors offer an alternative platform for constructing topological quantum gates, as they host Majorana modes at lower-dimensional boundaries \cite{ref15,ref16,ref17,ref18,ref19,ref20,ref21,ref22,ref23,ref24,ref25,ref26,ref27}. For instance, second-order topological superconductors support Majorana zero modes localized at sample corners \cite{ref28,ref29,ref30}. Braiding of these Majorana corner modes can therefore be achieved via parameter tuning, without the need to manipulate superconducting vortices \cite{ref31,ref32,ref33,ref34,ref35}. Nevertheless, existing schemes for generating isolated Majorana corner modes still typically rely on Zeeman fields. Taken together, realizations and manipulations of two-dimensional Majorana zero modes have so far generally required magnetic fields, which tend to suppress superconducting pairing.

Altermagnets exhibit collinear compensated magnetic order with momentum-dependent exchange fields \cite{ref36,ref37,ref38}. Their vanishing net magnetic field avoids Pauli pair breaking of superconductivity \cite{ref39}. Altermagnetism breaks time-reversal symmetry. Such symmetry breaking provides a prerequisite for isolated nondegenerate Majorana corner modes for topological quantum computation. Studies on square-lattice systems show the interplay between $d$-wave altermagnetism and $s$-wave superconductivity can host Majorana corner modes \cite{ref40,ref41}. These proposals exploit anisotropic effective mass on orthogonal edges originating from altermagnetic order. Other altermagnetic heterostructures also provide routes to Majorana zero modes \cite{ref42,ref43,ref44}. Majorana braiding has also been numerically simulated in patterned altermagnetic platforms \cite{ref45}. Variations in lattice symmetry modify mass projections along edges for other lattice geometries. The honeycomb lattice supports zigzag, armchair, and bearded edges.
Studies on honeycomb lattices have identified higher-order insulating phases in the Kane-Mele model \cite{ref46} and Majorana corner modes in altermagnet-superconductor heterostructures \cite{ref47}. However, the mechanism for altermagnetism-induced edge mass projections and Majorana braiding on honeycomb lattices remains unclear.

In this work, we show that Majorana corner modes can arise at corners formed by bearded or zigzag edges in the $d$-wave altermagnet–superconductor heterostructure.
The altermagnetic edge-mass projections depend on both edge orientation and termination, leading to distinct conditions for the emergence of corner modes.
Tuning the altermagnetic anisotropy and chemical potential then allows their positions to be adiabatically controlled.
This enables corner-pair selection, paired transport in a hexagon, and exchange of a single pair in a triangle.
We further extend this edge-mass control scheme to coupled-flake networks and construct the $X$, Hadamard, and CNOT gates within fixed-parity encodings.   

\section{Model Hamiltonian for the physical system}
\label{sec:model}

\begin{figure}[!htbp]
\centering
\setlength{\abovecaptionskip}{3pt}
\includegraphics[width=\columnwidth]{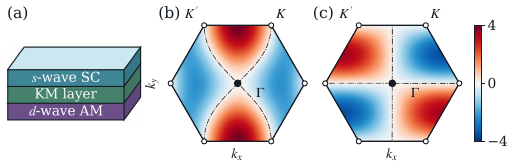}
\caption{(a) Illustration of a Kane--Mele layer sandwiched between an $s$-wave superconductor and a $d$-wave altermagnet.
(b)--(c) Altermagnetic form factor $f_{\AM}(\mathbf{k},\chi)/\lambda$ for $\chi=0$ ($d_{x^2-y^2}$) and $\chi=\pi/4$ ($d_{xy}$), respectively.
The hexagon denotes the first Brillouin zone. Dash-dotted lines mark the altermagnetic nodes.
Both panels adopt the same color scale.
}
\label{fig:model}
\end{figure}

Consider a Kane-Mele layer \cite{ref48} coupled to an $s$-wave superconductor and a
$d$-wave altermagnet as shown in Fig.~\ref{fig:model}(a). The Hamiltonian for the system reads
\begin{equation}
\begin{aligned}
H&=-t\sum_{\langle ij\rangle}c_i^\dagger c_j
+it'\sum_{\langle\!\langle ij\rangle\!\rangle}
\nu_{ij}c_i^\dagger s_zc_j+\sum_i(\varepsilon_i-\mu)c_i^\dagger c_i\\
&+\lambda\sum_{\langle\!\langle ij\rangle\!\rangle}
\eta_{ij}(\chi)c_i^\dagger(\boldsymbol{s}\cdot\hat{\boldsymbol n})c_j+\sum_i\bigl(
\Delta_sc_{i\uparrow}^\dagger c_{i\downarrow}^\dagger+
\mathrm{H.c.}\bigr),
\label{eq:KM}
\end{aligned}
\end{equation}
where $c_i=(c_{i\uparrow},c_{i\downarrow})^T$ denotes the annihilation operator spinor for spin-up and spin-down fermions at site $i$. \(\langle ij\rangle\) and \(\langle\langle ij\rangle\rangle\) label nearest-neighbor and next-nearest-neighbor site pairs, respectively. $t$ and $t'$ denote the nearest-neighbor hopping, intrinsic spin--orbit 
coupling, respectively. The chirality factor for the intrinsic spin--orbit coupling is
$\nu_{ij}=\operatorname{sgn}[\hat{\boldsymbol z}\cdot
(\hat{\boldsymbol d}_{ik}\times\hat{\boldsymbol d}_{jk})]=\pm1$.The unit vectors \(\hat{\boldsymbol{d}}_{ik}\) and \(\hat{\boldsymbol{d}}_{jk}\) point from the intermediate lattice site $k$ to sites $i$ and $j$. \(s_\alpha\) with \(\alpha=x,y,z\) denote the Pauli matrices acting on the spin degrees of freedom. \(\varepsilon_i=M\xi_i\) defines a staggered potential with \(\xi_i=1\) on the A sublattice and \(\xi_i=-1\) on the B sublattice, respectively.  
\(\lambda\) is the magnitude of the $d$-wave altermagnetic order, and \(\eta_{ij}(\chi)=\cos[2(\theta_{ij}-\chi)]\) encodes the $d$-wave anisotropy with $\theta_{ij}=\arg(\boldsymbol{r}_i-\boldsymbol{r}_j)$. Here \(\boldsymbol{r}_i\) and \(\boldsymbol{r}_j\) are the position vectors of sites $i$ and $j$, respectively. The unit vector \(\hat{\boldsymbol{n}}\) specifies the orientation of the Néel vector. In-plane Néel orientations are connected by global spin rotations about the \(\hat{\boldsymbol{z}}\) axis. We take the Néel vector \(\hat{\boldsymbol{n}}=\hat{\boldsymbol{y}}\) throughout this work unless otherwise specified. $\mu$ is the chemical potential and $\Delta_s$ the
$s$-wave pairing amplitude.

For the Bogoliubov–de Gennes (BdG) Hamiltonian of Eq.~\eqref{eq:KM}, its momentum-space kernel reads
\begin{equation}
h_{\BdG}(\boldsymbol k)=
\begin{pmatrix}
h(\boldsymbol k)&i\Delta_ss_y\\
-i\Delta_s^*s_y&-h^T(-\boldsymbol k)
\end{pmatrix},
\label{eq:BdG}
\end{equation}
under the basis $\Psi_{\boldsymbol k}=(C_{\boldsymbol k},C_{-\boldsymbol k}^\dagger)^T$,
with the spinor $C_{\boldsymbol k}=(c_{A\boldsymbol k\uparrow},c_{A\boldsymbol k\downarrow},c_{B\boldsymbol k\uparrow},c_{B\boldsymbol k\downarrow})$, where $h(\boldsymbol{k})=h_{\mathrm{KM}}(\boldsymbol{k})+f_{\AM}(\boldsymbol k,\chi)\sigma_0s_y+M\sigma_zs_0-\mu$. Here, \(\sigma_z\) acts on the sublattice degrees of freedom. \(\sigma_0\) and \(s_0\) denote the two-dimensional identity matrices.
The Kane--Mele Hamiltonian explicitly reads $h_{\mathrm{KM}}(\boldsymbol{k}) = b_{x0}\sigma_x s_0 + b_{y0}\sigma_y s_0 + b_{zz}\sigma_z s_z$,
where
$b_{x0}=-t\bigl[\cos k'_y + 2\cos k'_x \cos(k'_y/2)\bigr]$,
$b_{y0}=t\bigl[\sin k'_y - 2\cos k'_x \sin(k'_y/2)\bigr]$,
$b_{zz}=2t'\bigl[\sin 2k'_x - 2\sin k'_x \cos(3k'_y/2)\bigr]$ with $k'_x=k_x/2$ and $k'_y=k_y/\sqrt{3}$.
The altermagnetism form factor in momentum space reads
\begin{equation}
f_{\text{AM}}(\boldsymbol{k},\chi)=2\lambda\sum_{n=1}^3
\cos\big[2(\theta_n-\chi)\big]\cos(\boldsymbol{k}\cdot\boldsymbol{a}_n),
\label{eq:formfactor}
\end{equation}
where \(\boldsymbol{a}_1=(1,0)\) and \(\boldsymbol{a}_{2,3}=(1/2,\pm\sqrt{3}/2)\), with \(\theta_n=\arg\boldsymbol{a}_n\). The choices \(\chi=0\) and \(\chi=\pi/4\) yield the \(d_{x^2-y^2}\) and \(d_{xy}\) textures shown in Figs.~\ref{fig:model}(b) and \ref{fig:model}(c), respectively.

The honeycomb flake has bearded edges along $0^\circ$ and
$\pm60^\circ$, with alternating A and B terminations, as illustrated in Fig.~\ref{fig:phase}(b).
The upper (lower) horizontal edge terminates on B (A).
We also consider a bearded triangle with all three edges terminating on A.
The lowest-energy states of this bearded hexagon exhibit three distinct localization patterns as the altermagnetic strength \(\lambda\) increases as shown in Fig.~\ref{fig:phase}: extended along horizontal edges, localized at four corners, and paired at the two lateral vertices.
The two corner-localized phases host four near-zero-energy levels, with protection mechanisms distinct from those of the four-corner localized phase.

\FloatBarrier
\section{Low-energy edge theory and the origin of Majorana corner modes}
\label{sec:boundary-theory}

\begin{figure}[t!]
\centering
\includegraphics[width=\columnwidth]{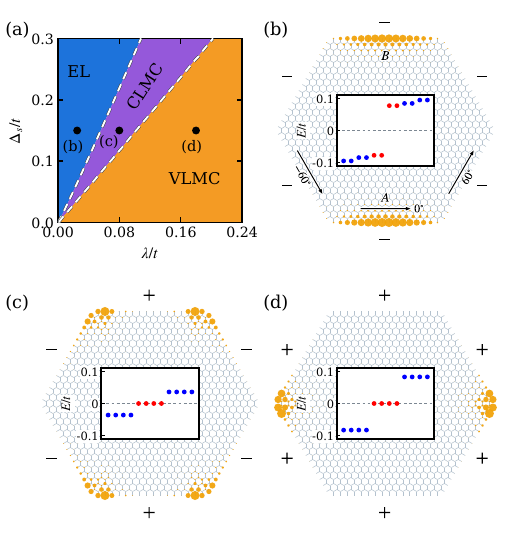}
\caption{(a) Phase diagram. EL denotes the edge-localized phase, CLMC denotes the corner-localized Majorana phase, and VLMC denotes the vertex-localized Majorana phase.
The EL, CLMC, and VLMC phases are separated by the phase boundaries $|\Delta_s|=3\lambda$ and $|\Delta_s|=3\lambda/2$.
(b) Density distributions of edge bound states (red dots in the inset) for a trivial superconductor. A and B denote A- and B-sublattice edge terminations.
(c)--(d) Density distributions of Majorana corner modes in the CLMC and VLMC phases.
The radii of the orange disks are proportional to the local density.
The insets show the energy levels. In (b)--(d), the parameters $|\Delta_s|$ and $\lambda$ are highlighted by black points in (a).
In all panels, $M=\mu=\chi=0$, and $t'=0.03t$.}
\label{fig:phase}
\end{figure}

At the bulk valleys, the $d$-wave altermagnetic form factor in Eq.~\eqref{eq:formfactor} cancels out. Edge projections depend on the boundary profile and sublattice termination \cite{ref30,ref49}. In the present model, this boundary dependence lifts the bulk cancellation and produces anisotropic edge masses.
For bearded edges, the nearest-neighbor zero mode at \(k_\parallel=0\) takes amplitudes \(\phi_n=(\sqrt{3}/2)(-1/2)^n\) on the terminating sublattice. Here \(n=0,1,\ldots\) labels rows moving inward from the boundary. We adopt this profile at leading order in \(t'/t\). The row overlaps read \(S_0=\sum_n\phi_n^2=1\) and \(S_1=\sum_n\phi_n\phi_{n+1}=-1/2\).
For an edge oriented at angle \(\theta_j\) relative to \(\boldsymbol{a}_1\), we define
\(\lambda_\parallel=\lambda\cos[2(\theta_j-\chi)]\) and
\(\lambda_\pm=\lambda\cos[2(\theta_j\pm\pi/3-\chi)]\).
The identity \(\lambda_+ + \lambda_- = -\lambda_\parallel\) then yields the projected contribution from the altermagnetic order
\begin{equation}
\begin{aligned}
m_j&=2\lambda_\parallel S_0+2(\lambda_++\lambda_-)S_1\\
&=3\lambda\cos[2(\theta_j-\chi)].
\end{aligned}
\label{eq:edgem}
\end{equation}
The negative adjacent-row overlap drives constructive addition between intra-row and inter-row contributions, yielding a finite edge gap when the sublattice potential \(u_j\) and superconductor pairing \(\Delta_s\) are neglected.

Projection of the staggered and chemical potentials yields
\(u_j=\xi_jM-\mu\), where \(\xi_j=+1\) (\(-1\)) labels A (B) termination. Onsite spin-singlet pairing maintains its amplitude because the unperturbed time-reversed edge-state partners carry unit orbital overlap.
For the uniform superconducting order \(\Delta_s=|\Delta_s|e^{i\phi}\) in Eq.~\eqref{eq:BdG}, the Nambu rotation
\(U_\phi=e^{i\phi\tau_z/2}\) satisfies
\(U_\phi^\dagger h_{\BdG}(\boldsymbol k,\phi)U_\phi
=h_{\BdG}(\boldsymbol k,0)\).
Therefore without loss of generality, we adopt this real-pairing gauge for all single-flake results below, unless stated otherwise.
Within the Nambu subspace of projected edge states, the edge Hamiltonian reads
\begin{equation}
h_{edge,j}(q)=v_0q s_z+m_js_y+u_j\tau_z-|\Delta_s|\tau_ys_y,
\label{eq:edgeH}
\end{equation}
where \(\tau_\alpha\) are Pauli matrices acting on Nambu space.
Boundary lengths are measured counterclockwise in lattice units.
\(v_0=-3t'\) denotes the Fermi velocity.
\(q= k- k_0\) is the small momentum, with \(k_0=0\) marking the gapless point of the edge states
when \(m_j\), \(u_j\), and \(\Delta_s\) vanish.
We retain terms linear in \(q\), assuming a gapped bulk
and an isolated edge doublet.

\begin{figure*}[t]
\centering
\includegraphics[width=\textwidth]{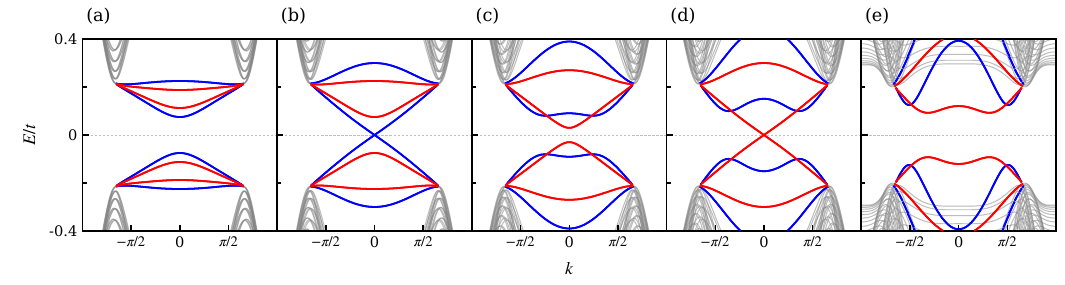}
\caption{Bearded-ribbon quasiparticle spectra at $M=\mu=\chi=0$ and $|\Delta_s|=0.15t$.
(a)--(e) $\lambda/t=0.025,0.050,0.080,0.100,0.180$, respectively.
Blue and red curves correspond to horizontal ($0^\circ$) and oblique ($\pm 60^\circ$) edge bands.
Gray curves represent the remaining bulk bands.
The spectrum for the $-60^\circ$ edge follows from the $60^\circ$ result under $k\mapsto -k$.
The horizontal-edge gap closes in panel (b), and the oblique-edge gap closes in panel (d).}
\label{fig:ribbons}
\end{figure*}

For the full edge Hamiltonian as in Eq.~\eqref{eq:edgeH}, its determinant takes the form
\begin{equation}
\det( h_{edge,j})=(v_0^2 q^2+m_j^2-R_j^2)^2+4|\Delta_s|^2 v_0^2 q^2,
\label{eq:det}
\end{equation}
where $R_j=\sqrt{u_j^2+|\Delta_s|^2}$.
For nonzero pairing, Eq.~\eqref{eq:det} vanishes
only at \(q=0\) and \(|m_j|=R_j\).
At this gap-closing point $q=0$, projection onto $s_y=\sigma=\pm1$ yields a Nambu doublet
$H_{j\sigma}(0)=\sigma m_j+u_j\tau_z-\sigma |\Delta_s|\tau_y,
\label{eq:edge-qzero}$
with eigenenergies $E_{\sigma,\eta}(0)=\sigma m_j+\eta R_j$ ($\eta=\pm1$).

The competition between altermagnetism and superconducting
pairing is captured by the effective edge mass
\begin{equation}
m_{\mathrm{edge},j}=|m_j|-\sqrt{u_j^2+|\Delta_s|^2},
\label{eq:closure}
\end{equation}
This quantity changes sign across the gap-closing transition. Its sign distinguishes the two distinct gapped edge phases (See Appendix~\ref{app:edge} for details).
The onsite potential $u_j$ modifies the critical threshold from $|\Delta_s|$ to $R_j$, whereas the global superconducting phase has no effect.
Its magnitude gives the minimum excitation energy at \(q=0\),
but need not equal the global gap.

Near a gap-closing point, projection onto the two critical states yields a Dirac Hamiltonian.
Matching neighboring edge bases at the corner and choosing positive velocity $v(\ell)$ gives its Hermitian real-space form,
\begin{equation}
h=-\frac{i}{2}\{v(\ell),\partial_\ell\}\widetilde\varrho_z
+m_{\mathrm{edge}}(\ell)\widetilde\varrho_y,
\label{eq:hermitian}
\end{equation}
where $\widetilde\varrho_\alpha$ are Pauli matrices acting on the critical doublet, and the corner lies at $\ell=0$.
Resolving $\widetilde\varrho_x=\beta=\pm1$, the zero-energy equation for the wave-function component $\psi_\beta$ reads
$\partial_\ell\!\left(\sqrt v\,\psi_\beta\right)
=-\beta\frac{m_{\mathrm{edge}}}{v}
\left(\sqrt v\,\psi_\beta\right)$.
Wave-function decay on both sides requires $\beta m_{\mathrm{edge}}(0^+)>0$ and
$\beta m_{\mathrm{edge}}(0^-)<0$.
The mass domain wall therefore hosts a Jackiw--Rebbi zero mode
\cite{ref25,ref50}.
Particle--hole symmetry allows its full Nambu spinor
to be chosen self-conjugate, identifying it as a Majorana mode.

For long gapped edges, we define $r_j=\operatorname{sgn}\bigl(m_{\mathrm{edge},j}\bigr)$.
The protected Majorana count is $N_{\mathrm{prot}}=\frac12\sum_{j=1}^{6}|r_{j+1}-r_j|$,
with $r_7=r_1$.
This formula counts protected single Majorana modes at each edge-phase interface.
Additional pairs can appear when an extra symmetry suppresses their hybridization.
The detailed projection and zero-mode envelope are given in
Appendix~\ref{app:edge}.
Overlap between well-separated corner modes induces exponentially small finite-energy splittings.

The orientation dependence of \(m_j\) yields distinct gap-closing conditions for horizontal and oblique ribbons as in Fig.~\ref{fig:ribbons}. For \(M=\mu=\chi=0\), the edge masses read
\(m_{\mathrm{edge},h}=3|\lambda|-|\Delta_s|\) and
\(m_{\mathrm{edge},o}=3|\lambda|/2-|\Delta_s|\).
When \(|\Delta_s|>3|\lambda|\), both masses are negative. No mass domain wall emerges in this regime, and the lowest-energy states carry finite energy with spectral weight concentrated on horizontal edges, where the edge gap is smaller as in Fig.~\ref{fig:phase}(b).
Upon increasing \(|\lambda|\), the horizontal edge mass changes sign first, while the oblique mass remains negative. This defines the four-corner phase window
\begin{equation}
\frac{1}{3}|\Delta_s|<|\lambda|<\frac{2}{3}|\Delta_s|.
\label{eq:four-corner-window}
\end{equation}
A Majorana mode localizes at each of the four horizontal endpoints within this window [Fig.~\ref{fig:phase}(c)]. At \(|\Delta_s|=0.15t\), our low-energy theory predicts gap closings at
\(\lambda=0.05t\) and \(0.10t\), consistent with the numerical results in
Figs.~\ref{fig:ribbons}(b) and \ref{fig:ribbons}(d).
These two thresholds delineate the four-corner region in the phase diagram Fig.~\ref{fig:phase}(a).

For \(|\Delta_s|<3|\lambda|/2\), both edge masses become positive and \(N_{\mathrm{prot}}=0\). Nevertheless, four near-zero-energy states persist in this phase, as seen in Fig.~\ref{fig:phase}(d).
At \(M=\mu=0\), the lattice Hamiltonian commutes with \(Q=\tau_y\sigma_z\) in the real-pairing gauge. \(\sigma_z\) acts on the sublattice degrees of freedom. In the edge theory, each lateral vertex hosts a pair
of Majorana modes in different $Q$ sectors, whose
mixing is forbidden by $Q$ conservation
(see Appendix~\ref{app:edge}). The term \(-\mu\tau_z\) couples the two sectors and lifts the degeneracy of each vertex mode pair without closing the edge gap. These vertex-bound mode pairs are therefore qualitatively distinct from the four isolated domain-wall Majorana corner modes shown in Fig.~\ref{fig:phase}(c).

Changing the edge termination modifies the boundary thresholds.
For zigzag edges, an analogous mass criterion as in Eq. (\ref{eq:closure}) holds, with
$m_j^{(z)}=-2p\lambda\cos[2(\theta_j-\chi)]$ and
$u_j^{(z)}=\xi_jpM-\mu$, where
$p=t/\sqrt{t^2+16t'^2}$. The pairing amplitude remains unaltered.
These coefficients encode the finite spin-orbit-coupling dependence inherited from the unperturbed edge states.
At $M=\mu=\chi=0$, the four-corner phase window reads
$|\Delta_s|/(2p)<|\lambda|<|\Delta_s|/p$. The derivation of this projected boundary condition is provided in Appendix~\ref{app:edge}.

For bearded-edge flakes, opposite edges satisfy $m_{j+3}=m_j$ while terminating on opposite sublattices. Their transition thresholds obey $R_{\mathrm{B}}^2-R_{\mathrm{A}}^2=4M\mu$, where $R_{\mathrm{A},\mathrm{B}}$ correspond to $R_j$ for $\mathrm{A}$-type and $\mathrm{B}$-type edge terminations.
Nonzero $M$ and $\mu$ applied simultaneously can split the transition points on opposite edges. This enables selective tuning of corner modes (See Appendix~\ref{app:paths}), which underpins the braiding protocols of Majorana corner modes.


\section{Tunable Majorana corner modes and Clifford quantum gates}
\label{sec:gates}

Sign reversal of an edge mass transfers a Majorana
mode between its endpoints, provided the two adjacent
edges carry opposite mass signs.
The onsite potentials set the threshold for this reversal,
while the altermagnetic texture determines the exchange
projection on each edge.
Together, these controls select which edge undergoes
a transition and thereby which pair of corners is connected
by the transfer.
The boundary criterion in Eq.~\eqref{eq:closure} thus provides a control
principle for routing Majorana corner modes.

We first examine the transfer and exchange of an isolated
Majorana pair on a single triangular flake, then extend
the analysis to interconnected flakes that provide the
additional modes and transfer branches required for
encoded Clifford gates.

For such connected device networks, we adopt
\begin{equation}
\begin{aligned}
H_{\mathrm{net}}(s)=\sum_a H_a(s)
-t_c\!\sum_{(ai,bj)\in\mathcal L}
\left(c_{ai}^{\dagger}c_{bj}+\mathrm{H.c.}\right),
\end{aligned}
\label{eq:network}
\end{equation}
where $H_a$ denotes the island Hamiltonian given in Eq.~\eqref{eq:KM},
$c_{ai}$ is the electron spinor, and $\mathcal L$ contains each
spin-conserving inter-island bond exactly once. The tunnel amplitude is real and fixed at $t_c=0.30t$. The parameter $s\in[0,1]$ parametrizes the control
trajectory $\{\chi_a(s),M_a(s),\mu_a(s)\}_a$,
with physical time $Ts$.
Each island supports uniform $s$-wave superconducting
pairing $\Delta_{s,a}=|\Delta_s|e^{i\phi_a}$ with fixed $\phi_a$.
Adopting the real-pairing gauge independently on each island yields
\begin{equation}
\widetilde T_{ab}
=-t_c\tau_z e^{i(\phi_b-\phi_a)\tau_z/2},
\label{eq:network-phase}
\end{equation}
with the spin identity matrix omitted. Fixed pairing-phase differences enter the inter-island tunnel coupling, whereas edge masses govern the mode-transfer pathways on each individual flake. Taken together, the low-energy edge theory and the full network BdG Hamiltonian describe the propagation of corner modes across these interconnected devices, as detailed in the subsequent subsections.

\subsection{Braiding of two Majorana corner modes}
\label{sec:exchange}

The triangular geometry has three edges, each associated with a distinct altermagnetic angle $\theta_{j}$.
Successive sign reversals of edge mass route two modes through the unoccupied vertex \cite{ref31}.
For an $\mathrm{A}$-terminated bearded triangular flake, all edges share the same threshold $R=\sqrt{(M-\mu)^2+|\Delta_s|^2}$.
The three altermagnetic induced edge masses read $3|\lambda\cos[2(\theta_j-\chi)]|$ with $\theta_{j=1,2,3}=0,\pi/3,2\pi/3$. The smallest never exceeds $3|\lambda|/2$, and the largest is at least $3\sqrt{3}|\lambda|/2$.
The inequality
\begin{equation}
\frac{3}{2}|\lambda|<\sqrt{(M-\mu)^2+|\Delta_s|^2}
<\frac{3\sqrt{3}}{2}|\lambda|
\label{eq:triangle-window}
\end{equation}
maintains mixed edge-mass signs for all noncritical orientations. This condition produces two mass domain walls.
At each reversal event, one mode spreads along the reversing edge, and then relocates to the opposite endpoint.

\begin{figure}[!tbp]
\centering
\includegraphics[width=\columnwidth]{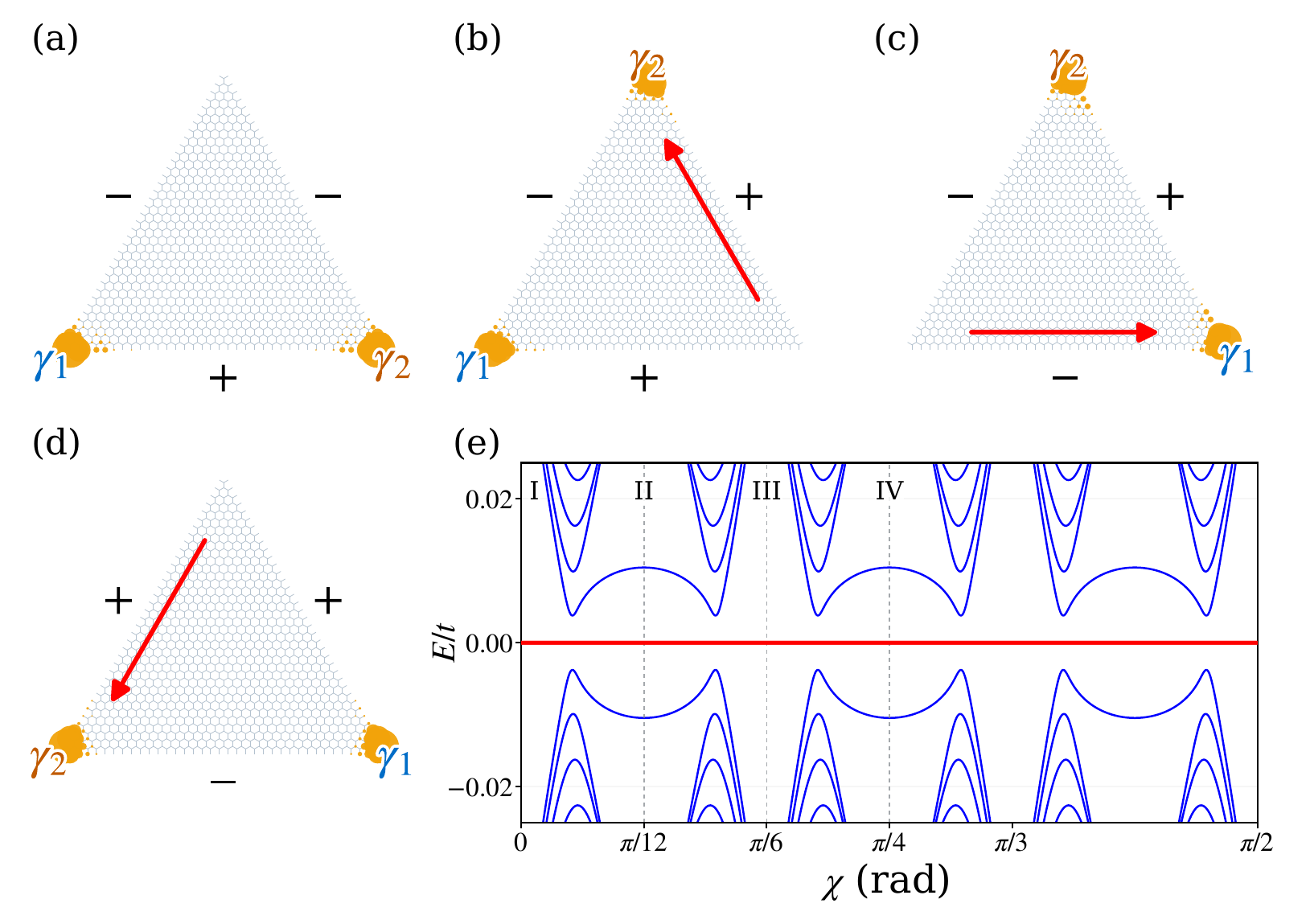}
\caption{Braiding of two Majorana corner modes in an $A$-terminated bearded triangle flake with 1404 lattice sites.
(a)--(d) Density distributions of Majorana corner modes at $\chi=0,\pi/12,\pi/6,\pi/4$, respectively.
Red arrows indicate the transfer directions of these Majorana corner modes.
The $+$ and $-$ signs denote positive and negative effective edge masses $m_{\mathrm{edge},j}$, respectively.
(e) Low-energy BdG spectrum for $0\leq\chi\leq\pi/2$.
Red curves mark the two Majorana zero-energy levels.
Guidelines I--IV mark the $\chi$ values used in panels (a)--(d).
The common parameters are $\lambda=0.09t$, $|\Delta_s|=0.15t$, $\mu=0.10t$, $M=0$, and $t'=0.03t$.}
\label{fig:triangle_exchange}
\end{figure}

For the parameters in Fig.~\ref{fig:triangle_exchange}, the bottom-edge mass starts positive, while the two oblique-edge masses start negative.
Upon increasing $\chi$, the right-edge mass reverses first, followed by the bottom-edge mass and then the left-edge mass.
The first reversal shifts $\gamma_2$ from the lower-right corner to the top vertex, with $\gamma_1$ remaining localized at the lower-left corner.
The second reversal transports $\gamma_1$ to the vacant lower-right corner. The third reversal moves $\gamma_2$ to the lower-left corner.
The two Majorana modes swap their positions at $\chi=\pi/4$. This sequence is illustrated in Figs.~\ref{fig:triangle_exchange}(a)--(d).
Along this control path, the spectrum of the finite triangular flake hosts an isolated near-zero-energy Majorana pair as shown in Fig.~\ref{fig:triangle_exchange}(e).
Edge delocalization during each mass-sign reversal establishes the spatial connection between the initial and final corners.

At $\chi=\pi/4$, the two modes occupy one another’s initial corners, and all three edge masses undergo sign reversal.
Proceeding to $\chi=\pi$ completes the texture period and recovers the Hamiltonian for a fixed N\'eel vector.
Parallel transport across this full cycle yields
\begin{equation}
\gamma_1\mapsto-\gamma_1,\quad\gamma_2\mapsto-\gamma_2.
\label{eq:triangle-cycle}
\end{equation}
See Appendix~\ref{app:transport} for more details.
This overall sign change arises from the geometric action associated with the full texture cycle.
The four spatial interchanges partition the parameter path into successive gapped segments.
Introducing a second Majorana pair enlarges the fixed-parity sector to form a two-dimensional logical subspace.
Braiding within this subspace generates qubit rotations.
Transport of paired Majorana corner modes in a bearded hexagonal flake is also demonstrated in Appendix~\ref{app:paths}.


\subsection{X gate}
\label{sec:Xgate}

In this subsection, we present an example demonstrating that Majorana braiding can also be implemented on hexagonal flakes to realize the X gate. The two coupled hexagonal flakes in Fig.~5(a) host four
Majorana corner modes, initially localized at the
horizontal-edge endpoints of island L, with island R
serving as an auxiliary transfer branch.
We define the complex fermions
\begin{equation}
\begin{gathered}
f_1=\frac{\gamma_1+i\gamma_2}{2},\qquad
f_2=\frac{\gamma_3+i\gamma_4}{2},
\end{gathered}
\label{eq:fermion-def}
\end{equation}
with occupation-number operators $n_j=f_j^\dagger f_j$.
The logical qubit is encoded in the even-parity sector as
\begin{equation}
\begin{gathered}
|0\rangle_{\mathrm L}=|00\rangle,\qquad
|1\rangle_{\mathrm L}=|11\rangle,
\end{gathered}
\label{eq:Xencoding}
\end{equation}
where $|11\rangle=f_1^\dagger f_2^\dagger|00\rangle$, and $\mathrm L$ denotes the logical basis.
We adopt $U_{ij}=\exp(\pi\gamma_i\gamma_j/4)$ for an oriented braiding operator and $U_{ij}^\dagger$ for its inverse \cite{ref51}. A Majorana mode $\gamma_k$ transforms under the braiding operator $U_{ij}$ as $U_{ij}\gamma_k U_{ij}^\dagger$.
All braiding matrices are defined relative to a common initial Majorana frame prior to composition, where the rightmost factor acts first.
Generator indices mark fixed reference positions, and colors in the diagrams track the propagating modes along the control path. These generators specify the ideal logical transformation. Parallel transport quantifies its physical implementation along the device control trajectories.

Each hexagonal flake features zigzag termination along its two right oblique edges and bearded termination along the remaining edges. At the fixed texture angle $\chi=\pi/4$, the projection of the altermagnetic form factor along horizontal edges vanishes, and the horizontal edge masses remain negative. Mixed edge terminations yield distinct thresholds for edge-mass sign reversal on the oblique edges. Using Eq.~\eqref{eq:closure} and the parameter $p$ introduced in Sec.~\ref{sec:boundary-theory}, the thresholds for edge-mass sign reversal occur at
$|\xi_jpM_a-\mu_a|=0.203t$ for zigzag edges and
$|\xi_jM_a-\mu_a|=0.360t$ for bearded edges.
Tuning $M_a$ and $\mu_a$ crosses the thresholds of the right oblique edges while the left oblique masses stay positive. This selects $\gamma_2$ and $\gamma_3$ for transport and pins $\gamma_1$ and $\gamma_4$ at their original corners.

\begin{figure}[!tbp]
\centering
\includegraphics[width=\columnwidth]{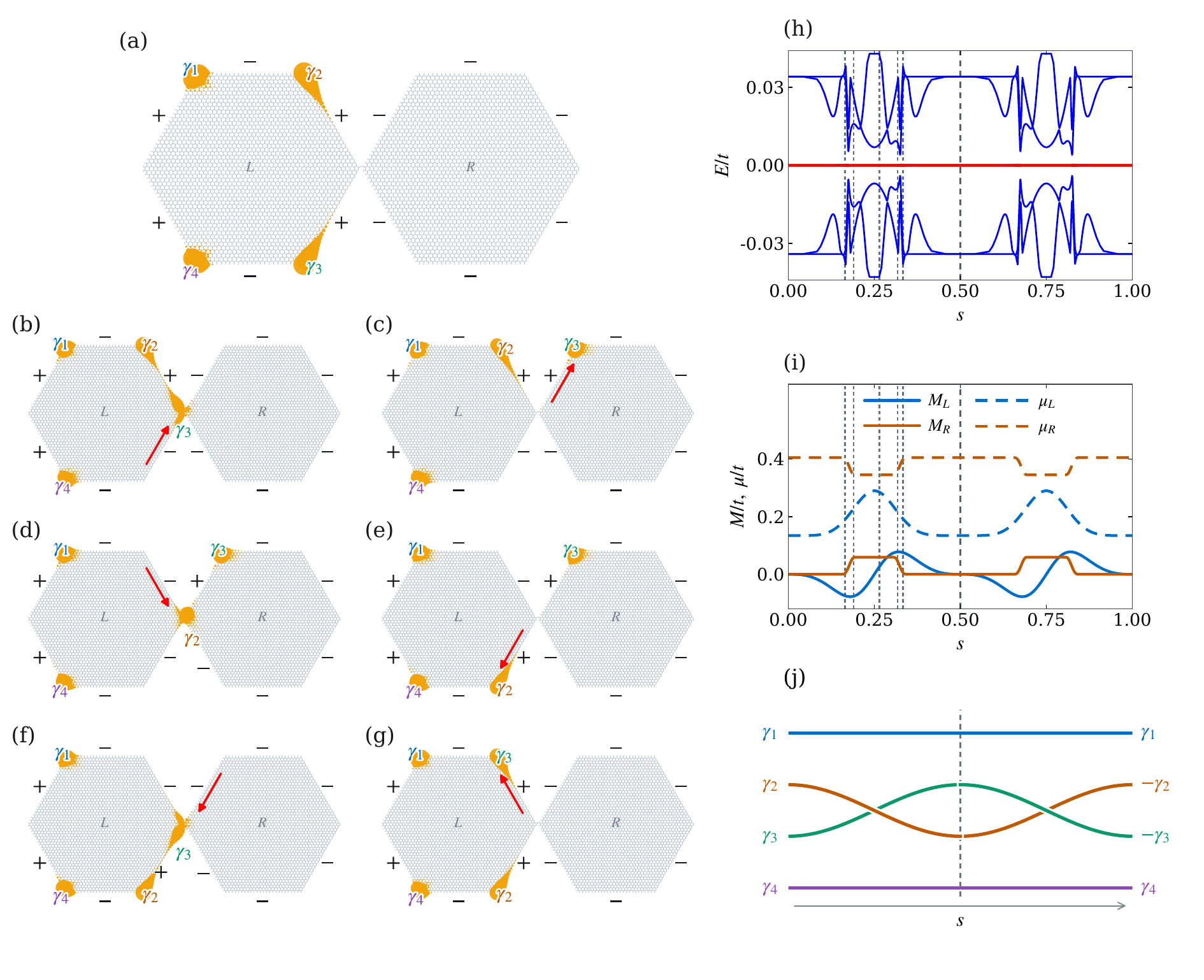}
\caption{$X$ gate via braiding in two coupled mixed-termination hexagonal flakes.
Each flake contains 3850 lattice sites.
The two flakes are connected by a single bond at their adjacent tips.
(a)--(g) Density distributions of Majorana modes during one braiding exchange, $0\leq s\leq 1/2$.
Labels track the Majorana modes.
Arrows indicate transfer directions.
The $+$ and $-$ signs denote positive and negative effective edge masses $m_{\mathrm{edge},j}$, respectively.
(h) Low-energy BdG energies along the full two-exchange braiding path.
Red curves mark Majorana zero mode levels and blue curves mark the remaining energy levels.
Dotted guidelines locate the snapshots (a)--(g).
(i) Onsite controls along the same path.
Solid and dashed curves denote controls $M_a$ and $\mu_a$.
(j) Braiding sequence along the full two-exchange path.
Endpoint signs are defined with respect to the initial reference frame.
Each half repeats the same closed control loop.
The parameters are $t'=0.06t$, $\lambda=|\Delta_s|=0.15t$, $\chi_L=\chi_R=\pi/4$, $t_c=0.30t$ and pairing phases $(\phi_L,\phi_R)=(0,\pi/3)$.
Initially, $M_L=M_R=0$, $\mu_L\simeq0.134t$, and $\mu_R\simeq0.405t$.}
\label{fig:Xgate}
\end{figure}

The auxiliary island permits one mode to be diverted from the main braiding path while the other proceeds onward. For $0\leq s\leq1/2$, we first flip the sign of the lower-right edge mass of island $L$, which brings $\gamma_3$ to the junction. A pulse applied to $M_R$, with $M_R+\mu_R$ kept constant, transfers $\gamma_3$ to the upper-left corner of island $R$, as shown in Figs.~\ref{fig:Xgate}(b) and (c). We then flip the sign of the upper-right edge mass of island $L$ to bring $\gamma_2$ toward the junction. Restoring the original sign of the lower-right edge mass moves $\gamma_2$ into the initial corner of $\gamma_3$, as depicted in Figs.~\ref{fig:Xgate}(d) and (e). Removing the pulse transports $\gamma_3$ back through the junction to the upper-right corner of island $L$, as illustrated in Figs.~\ref{fig:Xgate}(f) and (g). All control parameters return to their starting values at $s=1/2$, completing one braiding operation.

For this chosen orientation, the elementary braiding yields
\begin{equation}
U_{23}\gamma_2U_{23}^{\dagger}=-\gamma_3,\qquad
U_{23}\gamma_3U_{23}^{\dagger}=\gamma_2,
\label{eq:Xexchange}
\end{equation}
while $\gamma_1$ and $\gamma_4$ remain unchanged. Restricted to the even-parity subspace defined by Eq.~\eqref{eq:Xencoding},
$U_{23}|_{\mathrm{even}}=(I+iX)/\sqrt2$.
Repeating this braiding operation with identical orientation returns the Majoranas to their original corners while flipping the sign of $\gamma_2$ and $\gamma_3$.
Consequently, $f_1\mapsto f_1^\dagger$ and
$f_2\mapsto-f_2^\dagger$: the occupation of both complex fermions flips, while the total fermion parity remains unchanged. The resulting logical action reads
\begin{equation}
U_X|_{\mathrm{even}}=U_{23}^2|_{\mathrm{even}}=iX.
\label{eq:Xlogical}
\end{equation}
Two successive braiding operations map this geometrically induced sign transformation into a logical basis flip \cite{ref52}. The two stationary Majoranas provide the parity partners required for this gate action.

\par

\subsection{Hadamard gate}
\label{sec:Hgate}

\begin{figure}[!tbp]
\centering
\includegraphics[width=\columnwidth,trim=7bp 10bp 24bp 12bp,clip]{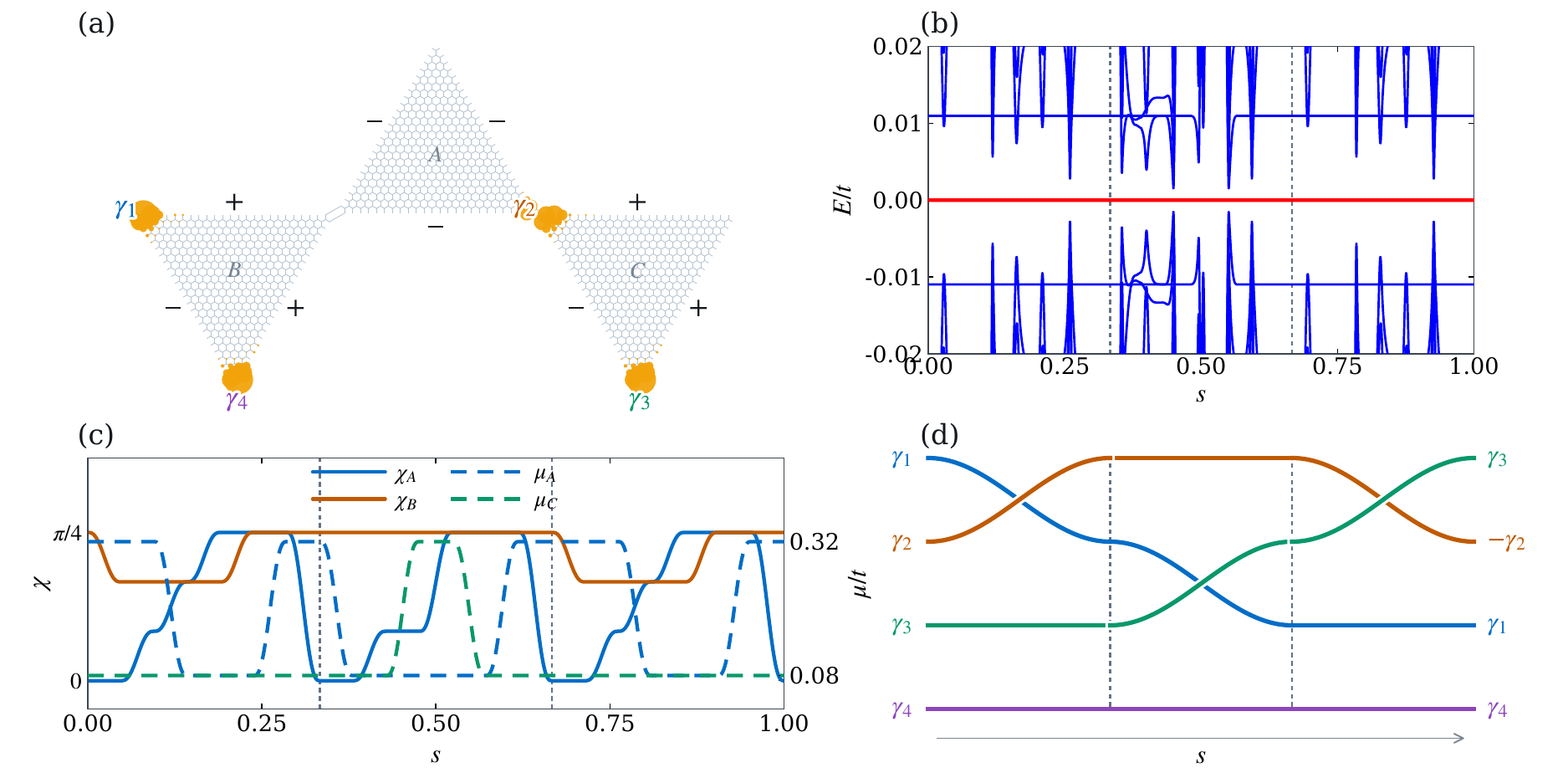}
\caption{Hadamard gate via braiding in three coupled bearded triangle flakes.
(a) Initial density distribution on the flakes with labels.
The signs $+$ and $-$ indicate positive and negative effective edge masses $m_{\mathrm{edge},j}$ in the initial configuration.
Each flake contains 648 lattice sites.
Each junction contains two interflake bonds with hopping strength $t_c$.
(b) Low-energy BdG energies along the closed transport path.
Energy curves follow the color convention of Fig.~\ref{fig:Xgate}(h).
(c) Control parameters along the same path.
Solid curves correspond to $\chi_A,\chi_B$.
Dashed curves correspond to $\mu_A,\mu_C$ on the right vertical scale.
(d) Braiding sequence.
Colors track the moving Majorana modes.
Rows mark the initial reference positions.
Endpoint signs are defined with respect to this reference frame.
The parameters are $t'=0.03t$, $\lambda=0.09t$, $|\Delta_s|=0.15t$, $M_a=0$, $t_c=0.30t$ and pairing phases $(\phi_A,\phi_B,\phi_C)=(0,2\pi/3,2\pi/3)$.
The fixed control parameters are $\chi_C=\pi/4$ and $\mu_B=0.08t$.
Initial values of all control parameters are listed in Appendix~\ref{app:transport}.}
\label{fig:Hgate}
\end{figure}

Access to a second logical rotation axis is gained by
exchanging the two Majorana modes of an encoded fermion,
complementing the rotation supplied by the X-gate protocol. This operation yields $U_{12}|_{\mathrm{even}}=e^{i\pi Z/4}$, with $Z|n\rangle_{\mathrm L}=(-1)^n|n\rangle_{\mathrm L}$.
The three-triangle flake network in Fig.~\ref{fig:Hgate}(a) makes both $U_{12}$ and $U_{23}$ accessible in the encoding of Eq.~\eqref{eq:Xencoding}. 
Central island $A$ provides a transfer pathway between islands $B$ and $C$. The Majorana zero mode at the lower tip of $B$ remains a spectator.

The control sequence combines texture rotations,
which modify the altermagnetic projection, with
chemical-potential pulses, which tune the competing
threshold $\sqrt{\mu_a^2+|\Delta_s|^2}$ at $M_a=0$.
Reducing $\mu_A$ supports domain-wall formation on the central island.
The first and third stages tune $\chi_A$ and $\chi_B$ to activate the $B$--$A$ branch. The intermediate stage fixes $\chi_B$ and applies a pulse on $\mu_C$ to open the transport route across the $A$--$C$ junction, as depicted in Fig.~\ref{fig:Hgate}(c). All control parameters return to their initial values after the three-stage sequence.
These three stages implement $U_{12}^{\dagger}$, $U_{23}^{\dagger}$, and $U_{12}^{\dagger}$ in chronological order, as depicted in Fig.~\ref{fig:Hgate}(d). These noncommuting logical rotations yield $U_H=U_{12}^{\dagger}U_{23}^{\dagger}U_{12}^{\dagger}$ with
\begin{equation}
\begin{aligned}
U_H|_{\mathrm{even}}&=-\frac{i}{\sqrt2}(X+Z)=-iH,
\end{aligned}
\label{eq:Hsequence}
\end{equation}
where $H=(X+Z)/\sqrt{2}$. Dropping the global U(1) phase factor, the logical-state transformation reads$|0\rangle_{\mathrm L}\mapsto \tfrac{1}{\sqrt{2}}\big(|0\rangle_{\mathrm L}+|1\rangle_{\mathrm L}\big),\quad
|1\rangle_{\mathrm L}\mapsto \tfrac{1}{\sqrt{2}}\big(|0\rangle_{\mathrm L}-|1\rangle_{\mathrm L}\big)$.
Successive rotations about the two distinct logical axes generate equal-amplitude superpositions, carrying the input-dependent relative sign characteristic of the Hadamard gate.

\par\FloatBarrier

\subsection{CNOT gate}
\label{sec:CNOTgate}
\suppressfloats[t]

\begin{figure}[!t]
\centering
\includegraphics[width=\columnwidth,trim=7bp 10bp 21bp 12bp,clip]{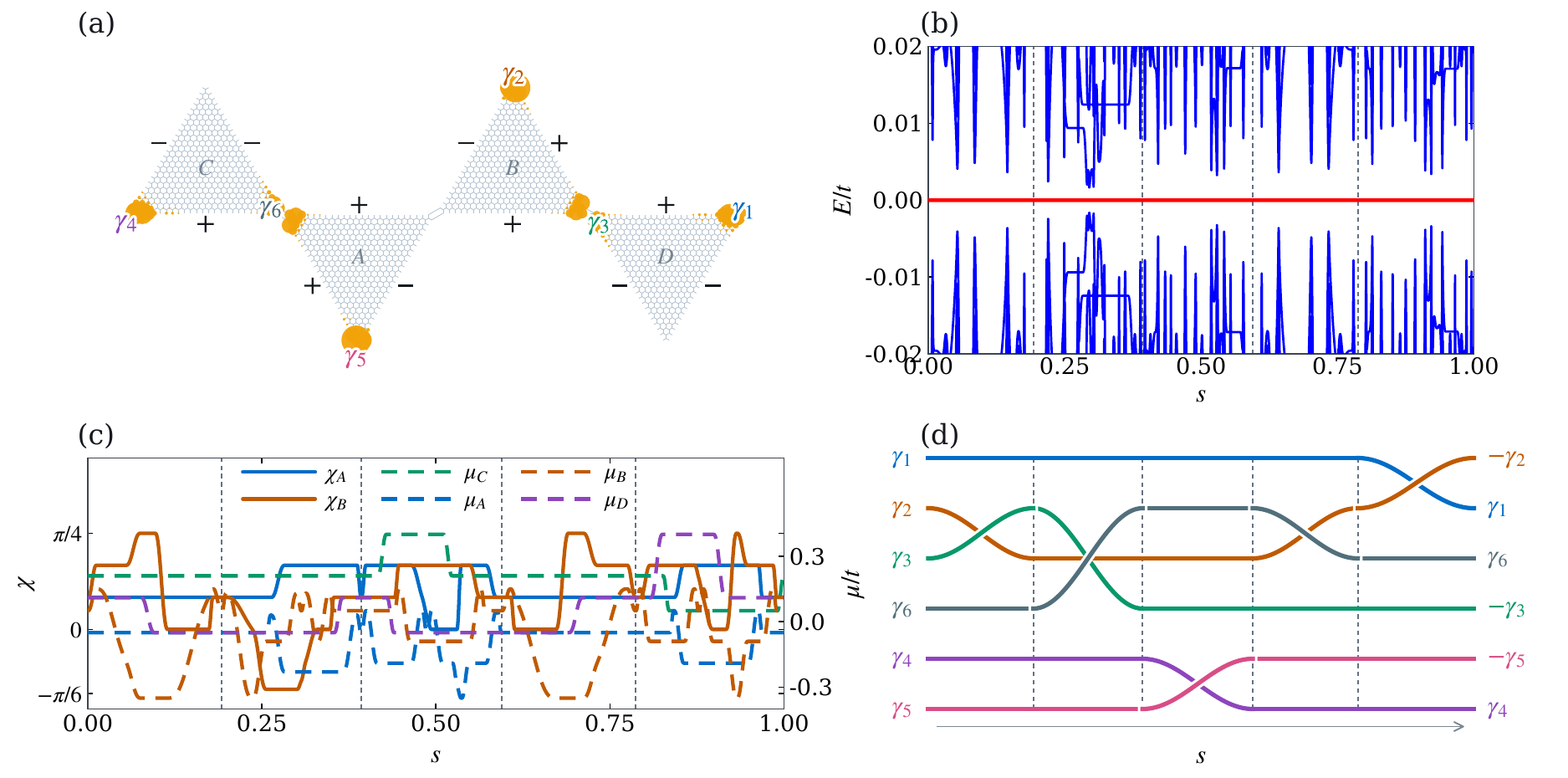}
\caption{CNOT gate via braiding in four coupled bearded triangle flakes.
These flakes share the same lattice size, $t'$, $\lambda$, $|\Delta_s|$, and interflake hopping bonds as in Fig.~\ref{fig:Hgate}.
(a) Initial density distributions of the six Majorana modes in the $C$--$A$--$B$--$D$ triangle flakes.
$+$ and $-$ denote the signs of the effective edge masses $m_{\mathrm{edge},j}$ in the initial configuration.
(b) Low-energy BdG energies.
Energy curves adopt the same color scheme as in Fig.~\ref{fig:Xgate}(h).
(c) Control parameters.
Solid curves represent $\chi_A,\chi_B$.
Dashed curves correspond to the four chemical potentials on the right vertical scale.
(d) Braiding sequence with five exchanges.
This panel adopts the same conventions as in Fig.~\ref{fig:Hgate}(d).
The row order is $1,2,3,6,4,5$.
The parameters are $M_a=0.05t$ and pairing phases $(\phi_C,\phi_A,\phi_B,\phi_D)=(0,2\pi/3,0,2\pi/3)$.
The fixed anisotropy textures are $\chi_C=\chi_D=\pi/30$.
Initial values of all control parameters are listed in Appendix~\ref{app:transport}.}
\label{fig:CNOTgate}
\end{figure}

To construct a conditional gate, we employ six Majorana zero modes within a common even-parity sector \cite{ref10}.
The four coupled triangle flakes $C$--$A$--$B$--$D$ in Fig.~\ref{fig:CNOTgate}(a) implement the encoding
\begin{equation}
\begin{gathered}
f_j=\frac{\gamma_{2j-1}+i\gamma_{2j}}{2},\qquad j=1,2,3,\\
| a,b\rangle_{\mathrm L}=| a,a\oplus b,b\rangle,\qquad a,b\in\{0,1\},
\end{gathered}
\label{eq:CNOTencoding}
\end{equation}
where the first line defines the three encoded fermion annihilation operators,
and the second line gives the logical basis.
The fermion creation operators follow the ordering $f_1^\dagger,f_2^\dagger,f_3^\dagger$.
Here $a$ labels the control bit and $b$ the target bit. $\oplus$ denotes addition modulo two.
This is a shared-parity encoding.
The middle occupation $n_2=a\oplus b$ correlates the two logical bits, while $n_1=a$ and $n_3=b$ remain independent.

The central $A$--$B$ flake link connects the two transfer branches $C$--$A$ and $B$--$D$.
Chemical-potential pulses on the outer islands and rotations of $\chi_A$, $\chi_B$ drive domain walls along these branches.
This sequence arranges the five braiding operations
$U_{23}^{\dagger}$, $U_{26}^{\dagger}$, $U_{45}^{\dagger}$, $U_{23}$, and $U_{12}^{\dagger}$ in chronological order.
After the first braiding, reference site $2$ hosts the mode originally labeled $\gamma_3$.
As a result, $U_{26}^{\dagger}$ corresponds to the crossing of the trajectories of $\gamma_3$ and $\gamma_6$, as depicted in Fig.~\ref{fig:CNOTgate}(d).
All control parameters return to their initial values after the fifth stage.
These ordered braiding operations yield
\begin{equation}
U_{\mathrm{CNOT}}=U_{12}^{\dagger}U_{23}U_{45}^{\dagger}
U_{26}^{\dagger}U_{23}^{\dagger}
\label{eq:CNOTsequence}
\end{equation}
within the common reference frame.
The induced fermion transformation reads
\begin{equation}
f_1\mapsto-if_1,\qquad f_2\mapsto if_3,\qquad f_3\mapsto if_2,
\label{eq:CNOTfermions}
\end{equation}
where these maps represent $U_{\mathrm{CNOT}}f_jU_{\mathrm{CNOT}}^\dagger$.
This transformation conserves $n_1$ and swaps $n_2$ and $n_3$.
Because the occupations $n_2$ and $n_3$ differ if and only
if $a=1$, their interchange flips the target bit $b$
precisely when the control bit is set.
Accounting for phase contributions from fermionic reordering yields
\begin{equation}
U_{\mathrm{CNOT}}|a,b\rangle_{\mathrm L}
=e^{-i\pi/4}| a,a\oplus b\rangle_{\mathrm L}.
\label{eq:CNOTaction}
\end{equation}
All four logical input states acquire the same global phase.
This produces the controlled bit flip while preserving the relative logical phases.
Up to this common phase, the gate maps $\tfrac{1}{\sqrt{2}}\big(|00\rangle_{\mathrm L}+|10\rangle_{\mathrm L}\big)$ to
$\tfrac{1}{\sqrt{2}}\big(|00\rangle_{\mathrm L}+|11\rangle_{\mathrm L}\big)$.
This transformation generates entanglement between the two encoded logical qubits.

The gapped energy spectra in Figs.~\ref{fig:Xgate}--\ref{fig:CNOTgate} preserve the selected four- or six-mode subspaces under the sampled control parameters.
Parallel transport with matched endpoints yields logical unitary matrices that closely approximate the target $X$, $H$, and CNOT gates, including their relative phases.
The phase-minimized Frobenius distances read
$\delta_X=5.65\times10^{-3}$, $\delta_H=2.04\times10^{-5}$,
and $\delta_{\mathrm{CNOT}}=1.87\times10^{-4}$.
Appendix~\ref{app:transport} details how the CNOT gate arises from composing the five braided transport matrices within the common reference frame.
This appendix also provides the energy scales and driving conditions required for coherent gate operation.
These control trajectories combine sign reversals of the edge mass to realize encoded single-qubit rotations and the conditional two-qubit gate.

\par\FloatBarrier

\section{Discussion and conclusion}

Directional altermagnetic order yields a finite edge projection despite canceling contributions at the bulk valleys.
On the honeycomb lattice, edge orientation and termination jointly determine this projection and the resulting edge masses.
Bearded and zigzag edges obey a similar edge-mass criterion but differ in their threshold values, which gives rise to Majorana corner modes.
Opposite edges of a hexagonal flake undergo simultaneous sign reversal of the edge mass when their thresholds coincide, while triangular flakes support successive mode transfers through a third vertex.
During transfer, the Majorana mode spreads along the edge whose mass reverses sign.
By organizing edge-mass reversals into closed control
cycles, we translate spatial Majorana transport into
geometric operations within fixed-parity encodings,
realizing the $X$, Hadamard, and CNOT gates with the
required relative logical phases.

The model platform is a Kane--Mele layer proximitized by an $s$-wave superconductor and a $d$-wave altermagnet.
The superconducting response of altermagnet--superconductor hybrids has been studied through their thermodynamic properties \cite{ref53} and finite-momentum Cooper pairing \cite{ref54}.
Separate material building blocks for this heterostructure have experimental precedents.
Bismuthene provides a promising honeycomb quantum spin Hall candidate \cite{ref55}, while monolayer WTe$_2$ shows signatures of proximity-induced pairing along its helical edge \cite{ref56}.
These systems separately validate the key ingredients of quantum spin Hall transport and superconducting proximity effect required for our heterostructure.
Reported altermagnetic band splitting \cite{ref57,ref58,ref59} and $d$-wave magnetic order in KV$_2$Se$_2$O \cite{ref60} suggest a viable material basis for directional altermagnetism, while domain manipulation in MnTe \cite{ref61} illustrates the possibility of tuning magnetic configurations.

Possible control mechanisms include superconducting phase bias, strain, and electrostatic gating.
Superconducting phase differences between neighboring islands can be set by phase bias, as demonstrated in phase-controlled Josephson junctions \cite{ref62}.
Strain engineering provides a possible route to tuning the altermagnetic anisotropy, as supported by theoretical studies of strained altermagnet--superconductor heterostructures \cite{ref63}.
Local electrostatic gates can tune the chemical potential $\mu$ \cite{ref64}, while a perpendicular electric field can generate a staggered potential $M$ in buckled honeycomb layers with sublattice-dependent electric-field coupling \cite{ref65}.
A functional device would require integration of these components into compatible heterointerfaces with suitable edge terminations while preserving proximity-induced pairing.

Taken together, our edge-mass control scheme provides a pathway for the manipulation of Majorana corner modes and the realization of Clifford gates in altermagnet–superconductor heterostructures.

\section{Acknowledgment}
Y. Gao and Y.-J. Wu was supported by NSFC under the grant No.12275203, Shaanxi Fundamental Science Research Project for Mathematics and Physics (Grant No.25JSY002), and the Scientific Research
Program Funded by Natural Science Foundation of Shaanxi Provincial Department of Education (Grant No. 23JK0483).

\appendix
\setlength{\parskip}{0pt}
\setlength{\textfloatsep}{12pt plus 2pt minus 2pt}
\setlength{\floatsep}{10pt plus 2pt minus 2pt}

\section{Edge projection and edge mass domain walls}
\label{app:edge}

\subsection{Edge projection}

We adopt the real-pairing gauge used in the main text.
The projection keeps terms linear in momentum and to
leading order in $\lambda$, $M$, $\mu$, and $|\Delta_s|$
within an isolated edge doublet.
For bearded edges, we use the leading-order profile in $t'/t$.
For zigzag edges, we retain the finite-$t'$ dependence
of the unperturbed edge states.

For a bearded edge parallel to $\boldsymbol a_1$,
the nearest-neighbor recursion
$t\phi_n+2t\cos(k/2)\phi_{n+1}=0$
gives $\rho_b(k)=-[2\cos(k/2)]^{-1}$,
where $n\geq0$ indexes rows inward from the boundary.
The normalized wavefunction profile at the crossing $k=0$ reads
\begin{equation}
|\phi_{j\sigma}\rangle
=\frac{\sqrt{3}}{2}
\sum_{n=0}^{\infty}
\left(-\frac{1}{2}\right)^n
|n,\xi_j,\sigma\rangle .
\label{eq:edgeprofile}
\end{equation}
Here $\sigma=\pm1$ labels the $s_z$ eigenstates and
$\xi_j=\pm1$ the terminating sublattice, on which the mode resides.
The same-row overlap equals unity and the adjacent-row
overlap equals $\rho_b$.
We define $\lambda_\parallel=\lambda\cos[2(\theta_j-\chi)]$
and $\lambda_\pm=\lambda\cos[2(\theta_j\pm\pi/3-\chi)]$,
which satisfy $\lambda_++\lambda_-=-\lambda_\parallel$.
The projected altermagnetic and intrinsic spin-orbit
contributions read
\begin{equation}
\begin{aligned}
h_{\mathrm{edge},\mathrm{AM}}
&=2\lambda_\parallel
  [\cos k-\rho_b\cos(k/2)]s_y,\\
h_{\mathrm{edge},\mathrm{SO}}
&=-2t'[\sin k-2\rho_b\sin(k/2)]s_z .
\end{aligned}
\label{eq:projection}
\end{equation}
The directed-bond convention for the $A$-terminated reference
edge is $(\nu_\parallel,\nu_+,\nu_-)=(1,-1,1)$.
At $k=0$, $\rho_b=-1/2$, giving $m_j=3\lambda_\parallel$
and the edge velocity $v_0=-3t'$.
Transforming the bond signs and momentum together gives
the same velocity for all bearded edges when momentum
is measured along the counterclockwise boundary direction.
The onsite potential projects to $u_j=\xi_jM-\mu$.
The unperturbed time-reversed edge-state partners have
unit orbital overlap, so onsite pairing retains its
amplitude $|\Delta_s|$.
These results give Eq.~\eqref{eq:edgeH} in the projected
Nambu basis
$(d_{jq\uparrow},d_{jq\downarrow},
d^\dagger_{j,-q,\uparrow},
d^\dagger_{j,-q,\downarrow})^T$.

For the zigzag edge, the band crossing occurs at $k=\pi$,
where finite spin-orbit coupling induces sublattice admixture.
Let $a=4t'/[\sqrt{t^2+16t'^2}+t]$.
On an $A$-terminated reference edge, the orbital amplitudes are
\begin{equation}
\begin{aligned}
\phi_{A,2n,\sigma}
&=\sqrt{1-a^2}(-a^2)^n,\\
\phi_{B,2n,\sigma}
&=\sigma a\,\phi_{A,2n,\sigma},
\end{aligned}
\label{eq:zigzag-profile}
\end{equation}
with zero amplitudes on odd rows.
The sublattice polarization and the opposite-spin orbital
overlap at the same momentum both equal
$p=(1-a^2)/(1+a^2)=t/\sqrt{t^2+16t'^2}$.
The adjacent-row altermagnetic matrix element vanishes
because odd rows carry no weight.
The same-row term therefore gives $-2p\lambda_\parallel$.
For the reference edge, the nearest-neighbor and intrinsic
spin-orbit terms contribute $4pt'$ and $2pt'$ to the
velocity, respectively.
The projected coefficients are
\begin{equation}
\begin{aligned}
m_j^{(z)}
&=-2p\lambda\cos[2(\theta_j-\chi)],\\
u_j^{(z)}
&=\xi_jpM-\mu,\qquad |v_j^{(z)}|=6pt'.
\end{aligned}
\label{eq:zigzag-coefficients}
\end{equation}
Onsite pairing links time-reversed edge states at $k=\pi$
and $k=-\pi$, whose orbital overlap remains unity.
Thus, $p$ renormalizes the altermagnetic contribution
and staggered potential, but not $|\Delta_s|$.

For bearded edges, with $R_j^2=u_j^2+|\Delta_s|^2$,
squaring Eq.~\eqref{eq:edgeH} gives
\begin{equation}
E_\pm^2
=v_0^2q^2+m_j^2+R_j^2
\pm2\sqrt{m_j^2R_j^2+u_j^2v_0^2q^2}.
\label{eq:spectrum}
\end{equation}
The product $E_+^2E_-^2$ recovers Eq.~\eqref{eq:det}
and the gap-closing condition in the main text.
The magnitude $|m_{\mathrm{edge},j}|$ gives the lowest
excitation energy at $q=0$, whereas the minimum over
momentum can occur at finite $q$ for $u_j\ne0$.
Within the projected Nambu basis,
$B_j=h_{\mathrm{edge},j}(0)\tau_x$ is antisymmetric,
with Pfaffian
\begin{equation}
\operatorname{Pf}B_j
=m_j^2-R_j^2
=(|m_j|+R_j)m_{\mathrm{edge},j}.
\label{eq:edge-pfaffian}
\end{equation}
With a common gapped high-momentum reference whose
Pfaffian sign is unchanged, opposite edge-mass signs
distinguish the two gapped edge phases~\cite{ref66,ref67}.
The same criterion applies to zigzag edges with
$m_j$ and $u_j$ replaced by $m_j^{(z)}$ and $u_j^{(z)}$.

\subsection{Mass domain walls and vertex modes}

Near a gap-closing point, the Nambu overlap between
the two critical states has magnitude $|\Delta_s|/R_j$.
Projection onto this critical doublet gives
\begin{equation}
h_{\mathrm{crit},j}(q)
=v_{\mathrm{crit},j}q\widetilde\varrho_z
+m_{\mathrm{edge},j}\widetilde\varrho_y,
\label{eq:critical}
\end{equation}
with $|v_{\mathrm{crit},j}|=|v_0\Delta_s|/R_j$.
In the matched continuum Hamiltonian of
Eq.~\eqref{eq:hermitian}, $v(\ell)>0$ approaches
$|v_{\mathrm{crit},j}|$ on each edge segment.
Integrating the zero-energy equation gives
\begin{equation}
\psi_\beta(\ell)\propto
\frac{|\widetilde\varrho_x=\beta\rangle}{\sqrt{v(\ell)}}
\exp\!\left[
-\beta\int_0^\ell
\frac{m_{\mathrm{edge}}(\ell')}{v(\ell')}
\,\mathrm d\ell'
\right],
\label{eq:zero}
\end{equation}
where $\beta=\pm1$.
Normalizability requires $\beta m_{\mathrm{edge}}>0$
on the $\ell>0$ edge and $\beta m_{\mathrm{edge}}<0$
on the $\ell<0$ edge, with masses evaluated away from
the corner.
The resulting domain-wall state can be chosen
self-conjugate under particle--hole symmetry,
$\mathcal C\psi=\psi$.
Its decay length near the gap closing is
$|v_{\mathrm{crit},j}/m_{\mathrm{edge},j}|$;
away from this limit, the decay exponents follow from
the full edge Hamiltonian.

For $M=\mu=0$, the lattice Hamiltonian commutes with
$Q=\tau_y\sigma_z$ and anticommutes with
$\mathcal S=\tau_y s_x$.
Particle--hole symmetry preserves the global $Q$-sector
label $\zeta=\pm1$.
On a bearded edge, $\tau_y=\xi_j\zeta$.
With $\varrho_x=\xi_j s_x$, $\varrho_y=s_y$, and
$\varrho_z=\xi_j s_z$, the sector Hamiltonian reads
\begin{equation}
h_{\mathrm{edge},j\zeta}(q)
=V_jq\varrho_z+\mathcal M_{j\zeta}\varrho_y,
\label{eq:sector}
\end{equation}
where $V_j=\xi_jv_0$ and
$\mathcal M_{j\zeta}=m_j-\xi_j\zeta|\Delta_s|$.
We assume corner matching that preserves $Q$,
$\mathcal S$, and probability current, with no additional
corner orbitals.
The zero-energy equation on each edge is then
$\partial_\ell\psi=-\kappa_{j\zeta}\varrho_x\psi$,
where $\kappa_{j\zeta}=\mathcal M_{j\zeta}/V_j
=(\xi_jm_j-\zeta|\Delta_s|)/v_0$.
Under this matching, each sector hosts a zero mode when
$\kappa_{j\zeta}\kappa_{j+1,\zeta}<0$.
The sign of $V_j$ enters the decay condition.
The solutions on adjacent edges are matched piecewise,
without interpolating $V_j$ through zero.

Numbering the edge segments counterclockwise from the
upper $B$-terminated horizontal edge gives $\xi_j=(-1)^j$.
At $\chi=0$, the quantities entering the sector decay
condition are
\begin{equation}
(\xi_1m_1,\ldots,\xi_6m_6)
=\frac{3}{2}\lambda(-2,-1,1,2,1,-1).
\label{eq:sector-pattern}
\end{equation}
In the window of Eq.~\eqref{eq:four-corner-window},
one sector wall occurs at each horizontal endpoint,
giving four isolated Majorana corner modes.
For $0<|\Delta_s|<3|\lambda|/2$, both sectors have walls
at junctions $2$--$3$ and $5$--$6$, giving two zero modes
at each lateral vertex.
These modes have opposite eigenvalues of
$\mathcal S=\zeta\varrho_x$ and belong to different
$Q$ sectors.
Conservation of $Q$ forbids their mixing.
A weak $-\mu\tau_z$ perturbation couples the sectors
and can split the pair while preserving $\mathcal S$
and the edge gap.
These additional pairs are not counted by $N_{\mathrm{prot}}$.

The continuum description requires decay lengths much
larger than the lattice spacing; well-separated corner
modes further require much longer edges.
For short decay lengths, quantitative splittings require
lattice eigenstates.
Finite-coupling corrections can shift the leading-order
transition thresholds.

\FloatBarrier

\renewcommand{\thefigure}{B\arabic{figure}}
\renewcommand{\theHfigure}{B.\arabic{figure}}
\setcounter{figure}{0}

\section{Majorana-mode localization and transport}
\label{app:paths}

\renewcommand{\thefigure}{B\arabic{figure}}
\renewcommand{\theHfigure}{B.\arabic{figure}}
\setcounter{figure}{0}

\subsection{Majorana-mode localization}

\begin{figure}[!t]
\centering
\includegraphics[width=0.92\columnwidth]{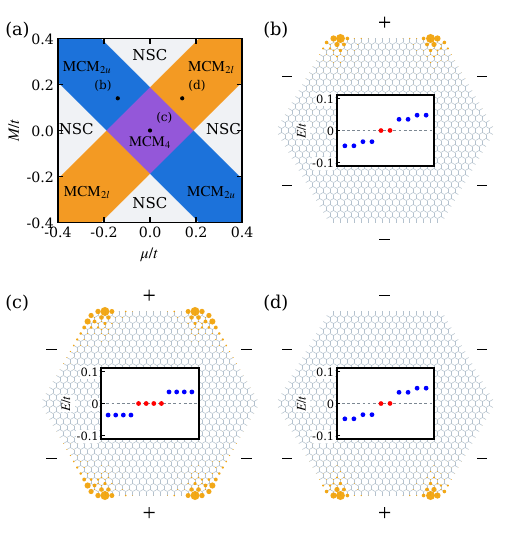}
\caption{(a) Phase diagram for the bearded hexagonal flake. NSC denotes the normal-superconductor phase with no Majorana corner modes. $\mathrm{MCM}_{2u}$ denotes the second-order topological superconducting phase hosting two Majorana corner modes localized at the ends of the upper horizontal edge. $\mathrm{MCM}_4$ denotes the second-order topological phase supporting four Majorana corner modes. $\mathrm{MCM}_{2l}$ denotes the second-order topological superconducting phase hosting two Majorana corner modes localized at the ends of the lower horizontal edge. Phase boundaries agree with the horizontal-edge criterion given in Eq.~\eqref{eq:pair-selection}. (b)–(d) present density distributions of the upper Majorana corner pair, four Majorana corner modes, and lower Majorana corner pair at $(M/t,\mu/t)=(0.14,-0.14)$, $(0,0)$, and $(0.14,0.14)$. These parameters are highlighted by black points in (a). Insets show the low-energy spectrum. Red dots mark the Majorana corner modes. Common parameters are at $\chi=0$,
 $t'=0.03t$, $\lambda=0.08t$, and $|\Delta_s|=0.15t$.}
\label{fig:selection}
\end{figure}

Alternating sublattice terminations allow selective
control of the upper and lower corner pairs in a bearded
hexagonal flake.
At $\chi=0$ and $3|\lambda|/2<|\Delta_s|<3|\lambda|$,
Eq.~\eqref{eq:closure} gives negative effective mass
on every oblique edge.
A positive mass on either horizontal edge therefore
produces a Majorana mode at each of its endpoints.
The upper and lower edges terminate on the $B$ and $A$
sublattices, with effective onsite potentials
$-M-\mu$ and $M-\mu$, respectively.
Their masses are positive when
\begin{equation}
\begin{aligned}
(M+\mu)^2+|\Delta_s|^2
&<9\lambda^2 \qquad (\mathrm{upper}),\\
(M-\mu)^2+|\Delta_s|^2
&<9\lambda^2 \qquad (\mathrm{lower}).
\end{aligned}
\label{eq:pair-selection}
\end{equation}
These inequalities define the overlapping strips in
Fig.~\ref{fig:selection}(a) within the leading-order
edge theory.
Their overlap hosts four Majorana corner modes;
within either strip outside the overlap, only the
corresponding pair remains.
Outside both strips, all edge masses are negative
and no mass domain walls occur.

Along $\mu=-M$, the effective onsite potentials vanish
on the upper edge and equal $2M$ on the lower edge.
Equation~\eqref{eq:pair-selection} thus retains only
the upper pair when $4M^2+|\Delta_s|^2>9\lambda^2$,
as illustrated by the finite-flake result in
Fig.~\ref{fig:selection}(b).
Along $\mu=M$, the upper and lower pairs exchange
roles [Fig.~\ref{fig:selection}(d)].
At $M=0$, the two thresholds coincide, so tuning $\mu$
removes both pairs simultaneously within the edge theory.

\subsection{Majorana-mode transport on a hexagonal flake}

Varying $\chi$ rotates the orbital altermagnetic texture
while keeping the N\'eel vector fixed.
The resulting edge-mass reversals move the domain walls,
with the gap-closing condition on edge $j$ given by
\begin{equation}
3\left|\lambda\cos\bigl[2(\theta_j-\chi)\bigr]\right|
=\sqrt{(\xi_jM-\mu)^2+|\Delta_s|^2}.
\label{eq:moving-wall}
\end{equation}
When the neighboring edges have opposite mass signs,
reversing the intervening edge mass transfers one
domain wall between its endpoints.
The mode spreads along that edge during the transition
and then relocalizes at the adjacent corner.

For $M=\mu=0$, opposite edges share the effective mass
$m_{\mathrm{edge},j}
=3|\lambda\cos[2(\theta_j-\chi)]|-|\Delta_s|$
and reverse their signs simultaneously.
Among the three edge orientations, the smallest
altermagnetic mass magnitude never exceeds
$3|\lambda|/2$, whereas the largest is at least
$3\sqrt{3}|\lambda|/2$.
Thus, positive and negative effective masses coexist
at every noncritical orientation when
\begin{equation}
\frac{3}{2}|\lambda|
<|\Delta_s|
<\frac{3\sqrt{3}}{2}|\lambda|.
\label{eq:paired-transport-window}
\end{equation}
The repeated mass pattern on opposite edges then
produces four domain walls, whose motion proceeds
through paired transfers.

\begin{figure}[!t]
\centering
\includegraphics[width=\columnwidth]{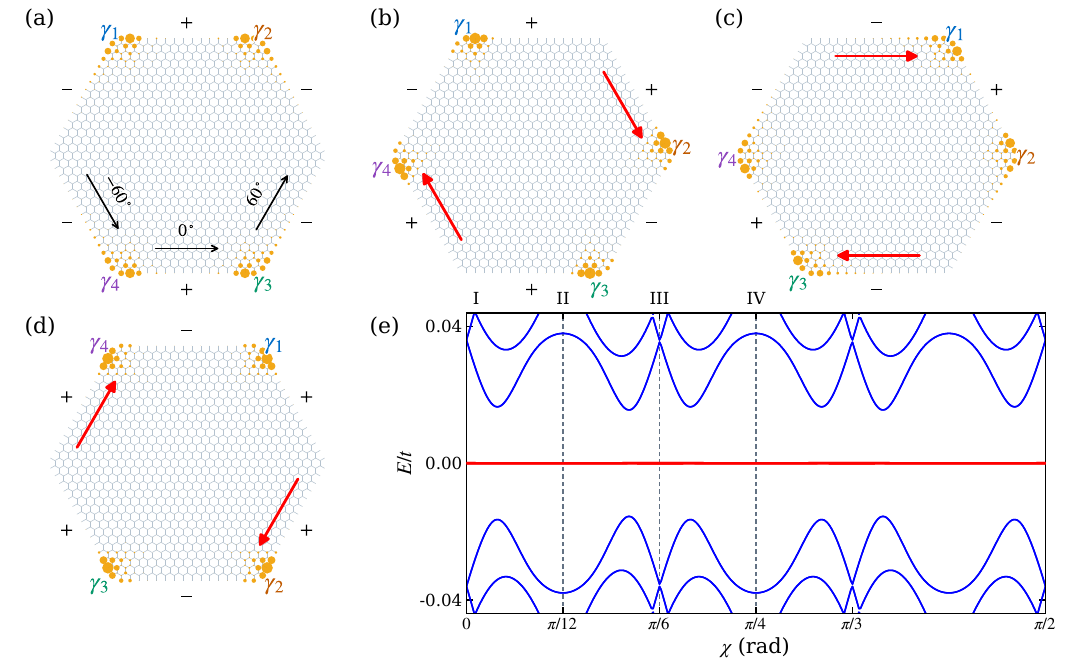}
\caption{Paired corner-mode transport in a bearded hexagon at $M=\mu=0$.
(a)--(d) Summed site probabilities of four Majorana modes at
$\chi=0,\pi/12,\pi/6,\pi/4$, with a common marker-size scale.
Colored labels track the modes; arrows show schematic transfer directions.
The $+$ and $-$ signs in (a)--(d) indicate the signs of the effective edge masses $m_{\mathrm{edge},j}$ at the corresponding values of $\chi$.
(e) Low-energy BdG spectrum. Red curves denote the four nearest-zero
levels and blue curves the other displayed levels. Guides I--IV
correspond to (a)--(d). The hexagon contains 1502 sites, with
$\lambda=0.08t$, $|\Delta_s|=0.15t$, and $t'=0.03t$.}
\label{fig:paired}
\end{figure}

Figure~\ref{fig:paired} illustrates this mechanism
for $\lambda=0.08t$ and $|\Delta_s|=0.15t$.
Define $\alpha=\tfrac12\arccos[|\Delta_s|/(3|\lambda|)]$,
which satisfies $\pi/12<\alpha<\pi/6$.
Equation~\eqref{eq:moving-wall} gives the first three
mass reversals at $\chi=\pi/6-\alpha$, $\alpha$,
and $\pi/3-\alpha$, respectively.
Starting from the four horizontal endpoints
[Fig.~\ref{fig:paired}(a)], $\gamma_2$ and $\gamma_4$
move to the lateral vertices when the upper-right and
lower-left edges reverse their masses
[Fig.~\ref{fig:paired}(b)].
The horizontal edges reverse next, transferring
$\gamma_1$ and $\gamma_3$ to their opposite endpoints
[Fig.~\ref{fig:paired}(c)].
The remaining oblique edges then reverse, carrying
$\gamma_2$ and $\gamma_4$ to the adjacent horizontal
endpoints.
By $\chi=\pi/4$, the original four corners are occupied
again, with the mode labels cyclically permuted
[Fig.~\ref{fig:paired}(d)].

The crossing angles above are predictions of the
leading-order edge theory.
The finite-flake spectrum in Fig.~\ref{fig:paired}(e)
shows four near-zero-energy states separated from the
other displayed levels at the sampled angles.
The effective edge-mass pattern has period $\pi/2$,
whereas the Hamiltonian with fixed N\'eel vector
has period $\pi$.
Consequently, the return to the original set of occupied
corners at $\chi=\pi/4$ represents a spatial permutation,
not a closed Hamiltonian cycle.

\section{Geometric evolution and finite-size effects}
\label{app:transport}

\subsection{Geometric evolution and logical gates}

The geometric action of the braiding paths in
Sec.~\ref{sec:gates} is obtained by transporting a
Majorana frame within the isolated low-energy subspace.
Let $P(s)$ project onto a particle--hole-invariant
$2N$-dimensional BdG subspace along the parameter curve
$\mathbf p(s)$, where $s\in[0,1]$ and the physical time
is $Ts$.
We choose an orthonormal frame satisfying
$\mathcal C\psi_a=\psi_a$; its vectors need not be
energy eigenstates.
The corresponding Majorana operators are
$\gamma_a=\sqrt{2}\psi_a^\dagger\hat\Psi$, where
$\hat\Psi$ is the real-space Nambu field, and satisfy
$\{\gamma_a,\gamma_b\}=2\delta_{ab}$.
The connection
$\mathcal A_{ab}(s)
=\langle\psi_a(s)|\partial_s\psi_b(s)\rangle$
is real and antisymmetric.

The single-flake Majorana pair corresponds to $N=1$.
The $X$ and $H$ gates use four-mode subspaces ($N=2$),
whereas the CNOT gate uses a six-mode subspace ($N=3$).
Writing $\Phi=(\psi_1,\ldots,\psi_{2N})$, the
parallel-transported frame is
$\Phi_{\mathrm{PT}}=\Phi W_{\mathrm{PT}}$, with
$\partial_sW_{\mathrm{PT}}=-\mathcal A W_{\mathrm{PT}}$
and $W_{\mathrm{PT}}(0)=I_{2N}$.
Endpoint matching for a closed parameter cycle gives
\begin{equation}
W=O_{\mathrm{end}}\mathcal P
\exp\left[-\int_0^1\mathcal A(s)\,\mathrm ds\right],
\label{eq:Wilson}
\end{equation}
where
$O_{\mathrm{end},ab}
=\langle\psi_a(0)|\psi_b(1)\rangle$
and $\mathcal P$ denotes path ordering.
For parity-preserving transport, $W\in\mathrm{SO}(2N)$.

Numerically, let $F_k$ be the aligned frame at the
$k$th sampling point and $\Phi_{k+1}$ the frame at
the next point.
We form the real overlap matrix
$O_k=\operatorname{Re}(\Phi_{k+1}^\dagger F_k)$
and take its singular value decomposition,
$O_k=L_k\Sigma_kR_k^{\mathsf T}$.
The next aligned frame is
$F_{k+1}=\Phi_{k+1}L_kR_k^{\mathsf T}$.
For a closed path, the transport matrix is then
$W=\operatorname{Re}(F_0^\dagger F_{\mathrm{end}})$.

For the single triangular flake, a spin rotation of
$\pi$ around the $z$ axis relates the Hamiltonians at
$\chi$ and $\chi+\pi/2$, which therefore have identical
spectra.
With the N\'eel vector fixed, the full parameter cycle
is $\chi:0\to\pi$.
Endpoint matching along this cycle gives $W=-I_2$,
consistent with Eq.~\eqref{eq:triangle-cycle}.

The associated many-body unitary operator $\hat U_W$
is defined in the initial frame by
$\hat U_W\gamma_a\hat U_W^\dagger
=\sum_bW_{ba}\gamma_b$.
For an oriented Majorana exchange, we write
$U_{ab}(\eta)=\exp(\eta\pi\gamma_a\gamma_b/4)$,
where $\eta=\pm1$ fixes the exchange orientation.
It satisfies
$U_{ab}\gamma_aU_{ab}^\dagger=-\eta\gamma_b$ and
$U_{ab}\gamma_bU_{ab}^\dagger=\eta\gamma_a$.
All exchange operators are expressed in the common
initial reference frame, with the rightmost factor
acting first.
Their indices refer to the initial Majorana labels.

Let $K$ be a real antisymmetric matrix such that $W=e^K$.
Its many-body lift reads
\begin{equation}
\hat U_W=\exp\!\left(
\frac14\sum_{a,b}K_{ab}\gamma_a\gamma_b
\right).
\label{eq:spin-lift}
\end{equation}
The two possible lifts differ by an overall sign.
The initial Majorana labels and the ordering of
creation operators follow Sec.~\ref{sec:gates}.
We restrict $\hat U_W$ to the logical basis
$\{|0\rangle_{\mathrm L},|1\rangle_{\mathrm L}\}$
for the single-qubit gates and
$\{|00\rangle_{\mathrm L},|01\rangle_{\mathrm L},
|10\rangle_{\mathrm L},|11\rangle_{\mathrm L}\}$
for CNOT.
The deviation of the resulting geometric operator
$U_G^{\mathrm{geom}}$ from the ideal gate $G$ is
\begin{equation}
\delta_G=\min_{\alpha\in\mathbb R}
\bigl\|U_G^{\mathrm{geom}}-e^{i\alpha}G\bigr\|_F,
\label{eq:gate-distance}
\end{equation}
where $\alpha$ is a real global phase angle,
$G=X,H,\mathrm{CNOT}$, and $\|\cdot\|_F$
denotes the Frobenius norm.
The minimization removes the overall phase while
retaining relative logical phases.
Thus, $\delta_G$ measures the geometric gate deviation,
not finite-time infidelity.

For the $X$ gate, the transport matrix of the closed
exchange is applied twice.
The $H$ gate is evaluated along the full
three-exchange path.
For CNOT, the five oriented exchanges in
Eq.~\eqref{eq:CNOTsequence} give
$W_{\mathrm{CNOT}}=W_5W_4W_3W_2W_1$,
where $W_k$ is the $k$th exchange matrix in
chronological order.
These matrices are evaluated in the common initial
six-mode frame; reversing a closed exchange replaces
its matrix by its transpose.
The control segments connect continuously and return
to the initial control configuration.

The initial Hadamard parameters, in island order
$(A,B,C)$, are
$(\chi_A,\chi_B,\chi_C)=(0,\pi/4,\pi/4)$,
$(\mu_A,\mu_B,\mu_C)=(0.32,0.08,0.08)t$,
and $M_a=0$.
For CNOT, they are
$\chi_A=\chi_B=\pi/12$,
$\chi_C=\chi_D=\pi/30$,
$(\mu_C,\mu_A,\mu_B,\mu_D)
=(0.21,-0.05,0.05,0.11)t$,
and $M_a=0.05t$.
The control paths are shown in
Figs.~\ref{fig:Hgate}(c) and \ref{fig:CNOTgate}(c).

The elementary closed exchange used for $X$ is
sampled adaptively, with 148 accepted points.
For $H$, the complete path contains 2011 points.
Its transport matrix differs from the coarser
three-exchange composition by
$8.93\times10^{-8}$ in Frobenius norm.
For CNOT, refining the sampling of the constituent
exchanges changes the composite transport matrix
by $3.51\times10^{-6}$.

\makeatletter
\@startsection{subsection}{2}{\z@}%
  {10pt}{7pt}%
  {\normalfont\small\bfseries\centering}%
  {Finite-size gaps and adiabatic conditions}
\makeatother

Let $e_{\mathrm{in}}(s)$ be the largest $|E|$ inside
$P(s)$ and $e_{\mathrm{out}}(s)$ the smallest $|E|$
outside it.
The spectral separation along the path is
$\Delta_{\mathrm{ex}}
=\min_s[e_{\mathrm{out}}(s)-e_{\mathrm{in}}(s)]$,
and the maximum local splitting ratio is
$r_{\max}
=\max_s[e_{\mathrm{in}}(s)/e_{\mathrm{out}}(s)]$.

\begin{table}[!t]
\centering
\caption{Spectral and geometric checks for the quantum gates.
Extrema are taken over the sampled control paths,
with local refinement for CNOT.
The geometric distance $\delta_G$ is defined in
Eq.~\eqref{eq:gate-distance}.}
\label{tab:gate-checks}
\setlength{\tabcolsep}{2.8pt}
\begin{tabular}{lcccc}
\hline\hline
Gate
& $\Delta_{\mathrm{ex}}/t$
& $\varepsilon_{\mathrm{split}}/t$
& $r_{\max}$
& $\delta_G$\\
\hline
X
& $3.91\times10^{-3}$
& $1.84\times10^{-4}$
& $3.43\times10^{-2}$
& $5.65\times10^{-3}$\\
H
& $1.54\times10^{-3}$
& $6.37\times10^{-7}$
& $5.81\times10^{-5}$
& $2.04\times10^{-5}$\\
CNOT
& $1.64\times10^{-3}$
& $5.69\times10^{-6}$
& $3.09\times10^{-4}$
& $1.87\times10^{-4}$\\
\hline\hline
\end{tabular}
\end{table}

Delocalization along an edge can increase the
hybridization between Majorana modes without closing
the spectral separation from the other states.
Up to an additive constant, the low-energy Hamiltonian
takes the form
\begin{equation}
H_{\mathrm{low}}
=\frac i2\sum_{a<b}\varepsilon_{ab}\gamma_a\gamma_b
+\sum_nE_nd_n^\dagger d_n,
\label{eq:lowH}
\end{equation}
where $\varepsilon$ is real and antisymmetric,
and $d_n$ annihilates an additional quasiparticle
excitation with energy $E_n>0$.
With this normalization,
$\|\varepsilon(s)\|_2=e_{\mathrm{in}}(s)$.
The maximum internal splitting is therefore
$\varepsilon_{\mathrm{split}}
=\max_s\|\varepsilon(s)\|_2
=\max_s e_{\mathrm{in}}(s)$.

Table~\ref{tab:gate-checks} lists these spectral
quantities and the geometric gate distances.
The CNOT path is locally refined near the spectral
bottlenecks.
All extrema are evaluated on the sampled paths and
do not constitute rigorous bounds along the
continuous curves.

A favorable operating hierarchy is
$\varepsilon_{\mathrm{split}}
\ll\hbar/T\ll\Delta_{\mathrm{ex}}$~\cite{ref68}.
The first inequality limits the accumulation of
internal dynamical phases~\cite{ref69}.
The ratios
$\varepsilon_{\mathrm{split}}/\Delta_{\mathrm{ex}}$
are $4.71\times10^{-2}$, $4.15\times10^{-4}$,
and $3.47\times10^{-3}$ for $X$, $H$, and CNOT,
respectively.
The separation of these energy scales is therefore
more limited for $X$.

Suppression of transitions out of the low-energy
subspace also depends on the control path and sweep
speed~\cite{ref70}.
For physical time $Ts$, the local adiabaticity
condition reads
\begin{equation}
\frac{\hbar}{T}\max_{s,a,n}
\frac{
|\langle n|\partial_sH_{\BdG}|a\rangle|
}{
|E_n-E_a|^2
}
\ll1,
\label{eq:adiabatic}
\end{equation}
where $|a\rangle$ and $|n\rangle$ are instantaneous
BdG eigenstates inside and outside $P(s)$, respectively.
The spectral hierarchy alone does not establish
this condition; its evaluation requires the
transition matrix elements along the chosen path.


\begin{thebibliography}{99}
\bibitem{ref1} C. Nayak, S. H. Simon, A. Stern, M. Freedman, and S. Das Sarma, Non-Abelian
anyons and topological quantum computation,
\href{https://doi.org/10.1103/revmodphys.80.1083}{Rev. Mod. Phys. \textbf{80}, 1083
(2008)}.

\bibitem{ref2} S. Das Sarma, M. Freedman, and C. Nayak, Majorana zero modes and
topological quantum computation,
\href{https://doi.org/10.1038/npjqi.2015.1}{npj Quantum Inf. \textbf{1}, 15001 (2015)}.

\bibitem{ref3} N. Read and D. Green, Paired states of fermions in two
dimensions with breaking of parity and time-reversal symmetries and the fractional
quantum Hall effect,
\href{https://doi.org/10.1103/PhysRevB.61.10267}{Phys. Rev. B \textbf{61}, 10267 (2000)}.

\bibitem{ref4} A. Yu. Kitaev, Unpaired Majorana fermions in quantum wires,
\href{https://doi.org/10.1070/1063-7869/44/10S/S29}{Phys. Usp. \textbf{44}, 131 (2001)}.

\bibitem{ref5} R. M. Lutchyn, E. P. A. M. Bakkers, L. P. Kouwenhoven, P. Krogstrup, C. M.
Marcus, and Y. Oreg, Majorana zero modes in superconductor--semiconductor
heterostructures,
\href{https://doi.org/10.1038/s41578-018-0003-1}{Nat. Rev. Mater. \textbf{3}, 52 (2018)}.

\bibitem{ref6} E. Prada, P. San-Jose, M. W. A. de Moor, A. Geresdi, E. J. H. Lee, J.
Klinovaja, D. Loss, J. Nyg\aa rd, R. Aguado, and L. P. Kouwenhoven, From Andreev to
Majorana bound states in hybrid superconductor--semiconductor nanowires,
\href{https://doi.org/10.1038/s42254-020-0228-y}{Nat. Rev. Phys. \textbf{2}, 575 (2020)}.

\bibitem{ref7} K. Flensberg, F. von Oppen, and A. Stern, Engineered platforms for
topological superconductivity and Majorana zero modes,
\href{https://doi.org/10.1038/s41578-021-00336-6}{Nat. Rev. Mater. \textbf{6}, 944
(2021)}.

\bibitem{ref8} P. Marra, Majorana nanowires for topological quantum computation,
\href{https://doi.org/10.1063/5.0102999}{J. Appl. Phys. \textbf{132}, 231101 (2022)}.

\bibitem{ref9} D. Aasen \emph{et~al.}, Blueprint for fault-tolerant quantum computation
with topological qubit arrays,
\href{https://doi.org/10.1103/qx36-4rv1}{Phys. Rev. Research \textbf{7}, 041002 (2025)}.

\bibitem{ref10} E. Mascot, T. Hodge, D. Crawford, J. Bedow, D. K. Morr, and S. Rachel,
Many-body Majorana braiding without an exponential Hilbert space,
\href{https://doi.org/10.1103/physrevlett.131.176601}{Phys. Rev. Lett. \textbf{131},
176601 (2023)}.

\bibitem{ref11} J. Bedow, E. Mascot, T. Hodge, S. Rachel, and D. K. Morr, Simulating
topological quantum gates in two-dimensional magnet-superconductor hybrid structures,
\href{https://doi.org/10.1038/s41535-024-00703-w}{npj Quantum Mater. \textbf{9}, 99
(2024)}.

\bibitem{ref12} C. Peeters, T. Hodge, E. Mascot, and S. Rachel, Effect of impurities and
disorder on the braiding dynamics of Majorana zero modes,
\href{https://doi.org/10.1103/physrevb.110.214506}{Phys. Rev. B \textbf{110}, 214506
(2024)}.

\bibitem{ref13} D. Crawford, R. Wiesendanger, and S. Rachel, Preparation and readout of
Majorana qubits in magnet-superconductor hybrid systems,
\href{https://doi.org/10.1103/physrevb.110.l220505}{Phys. Rev. B \textbf{110}, L220505
(2024)}.

\bibitem{ref14} T. Hodge, E. Mascot, D. Crawford, and S. Rachel, Characterizing dynamic
hybridization of Majorana zero modes for universal quantum computing,
\href{https://doi.org/10.1103/physrevlett.134.096601}{Phys. Rev. Lett. \textbf{134},
096601 (2025)}.

\bibitem{ref15} M. Geier, L. Trifunovic, M. Hoskam, and P. W. Brouwer, Second-order
topological insulators and superconductors with an order-two crystalline symmetry,
\href{https://doi.org/10.1103/physrevb.97.205135}{Phys. Rev. B \textbf{97}, 205135
(2018)}.

\bibitem{ref16} E. Khalaf, Higher-order topological insulators and superconductors
protected by inversion symmetry,
\href{https://doi.org/10.1103/physrevb.97.205136}{Phys. Rev. B \textbf{97}, 205136
(2018)}.

\bibitem{ref17} Y.-T. Hsu, W. S. Cole, R.-X. Zhang, and J. D. Sau, Inversion-protected
higher-order topological superconductivity in monolayer WTe$_2$,
\href{https://doi.org/10.1103/physrevlett.125.097001}{Phys. Rev. Lett. \textbf{125},
097001 (2020)}.

\bibitem{ref18} X.-H. Pan, K.-J. Yang, L. Chen, G. Xu, C.-X. Liu, and X. Liu,
Lattice-symmetry-assisted second-order topological superconductors and Majorana patterns,
\href{https://doi.org/10.1103/physrevlett.123.156801}{Phys. Rev. Lett. \textbf{123},
156801 (2019)}.

\bibitem{ref19} J. Ahn and B.-J. Yang, Higher-order topological superconductivity of
spin-polarized fermions,
\href{https://doi.org/10.1103/physrevresearch.2.012060}{Phys. Rev. Research \textbf{2},
012060(R) (2020)}.

\bibitem{ref20} S. Ikegaya, W. B. Rui, D. Manske, and A. P. Schnyder, Tunable Majorana
corner modes in noncentrosymmetric superconductors: Tunneling spectroscopy and edge
imperfections,
\href{https://doi.org/10.1103/physrevresearch.3.023007}{Phys. Rev. Research \textbf{3},
023007 (2021)}.

\bibitem{ref21} K. H. Wong, M. R. Hirsbrunner, J. Gliozzi, A. Malik, B. Bradlyn, T. L.
Hughes, and D. K. Morr, Higher order topological superconductivity in
magnet-superconductor hybrid systems,
\href{https://doi.org/10.1038/s41535-023-00564-9}{npj Quantum Mater. \textbf{8}, 31
(2023)}.

\bibitem{ref22} K. H. Wong, J. Gliozzi, M. R. Hirsbrunner, A. Malik, B. Bradlyn, T. L.
Hughes, and D. K. Morr, Competing higher order topological superconducting phases in
triangular lattice magnet-superconductor hybrid systems,
\href{https://doi.org/10.1103/physrevb.109.144521}{Phys. Rev. B \textbf{109}, 144521
(2024)}.

\bibitem{ref23} A. Chew, Y. Wang, B. A. Bernevig, and Z.-D. Song,
Higher-order topological superconductivity in twisted bilayer graphene,
\href{https://doi.org/10.1103/physrevb.107.094512}{Phys. Rev. B \textbf{107}, 094512 (2023)}.

\bibitem{ref24} P. M. Bonetti, D. Chakraborty, X. Wu, and A. P. Schnyder,
Interaction-driven first-order and higher-order topological superconductivity,
\href{https://doi.org/10.1103/physrevb.109.l180509}{Phys. Rev. B \textbf{109}, L180509 (2024)}.

\bibitem{ref25} Z. Yan, F. Song, and Z. Wang, Majorana corner modes in a high-temperature
platform,
\href{https://doi.org/10.1103/physrevlett.121.096803}{Phys. Rev. Lett. \textbf{121},
096803 (2018)}.

\bibitem{ref26} Y.-J. Wu, J. Hou, Y.-M. Li, X.-W. Luo, X. Shi, and C. Zhang, In-plane
Zeeman-field-induced Majorana corner and hinge modes in an $s$-wave superconductor
heterostructure,
\href{https://doi.org/10.1103/physrevlett.124.227001}{Phys. Rev. Lett. \textbf{124},
227001 (2020)}.

\bibitem{ref27} Y. Volpez, D. Loss, and J. Klinovaja, Second-order topological
superconductivity in $\pi$-junction Rashba layers,
\href{https://doi.org/10.1103/physrevlett.122.126402}{Phys. Rev. Lett. \textbf{122},
126402 (2019)}.

\bibitem{ref28} M. Kheirkhah, D. Zhu, J. Maciejko, and Z. Yan,
Corner- and sublattice-sensitive Majorana zero modes on the kagome lattice,
\href{https://doi.org/10.1103/physrevb.106.085420}{Phys. Rev. B \textbf{106}, 085420 (2022)}.

\bibitem{ref29} X.-J. Wang, Y. Mo, Z. Wang, Z. Wu, and Z. Yan,
Intrinsic second-order topological superconductors with tunable Majorana zero modes,
\href{https://doi.org/10.1103/sw5x-pqpr}{Phys. Rev. B \textbf{113}, 144513 (2026)}.

\bibitem{ref30} D. Zhu, M. Kheirkhah, and Z. Yan,
Sublattice-enriched tunability of bound states in second-order topological insulators
and superconductors,
\href{https://doi.org/10.1103/physrevb.107.085407}{Phys. Rev. B \textbf{107}, 085407 (2023)}.

\bibitem{ref31} S.-B. Zhang, W. B. Rui, A. Calzona, S.-J. Choi, A. P. Schnyder, and B.
Trauzettel, Topological and holonomic quantum computation based on second-order
topological superconductors,
\href{https://doi.org/10.1103/physrevresearch.2.043025}{Phys. Rev. Research \textbf{2},
043025 (2020)}.

\bibitem{ref32} T. E. Pahomi, M. Sigrist, and A. A. Soluyanov, Braiding Majorana corner
modes in a second-order topological superconductor,
\href{https://doi.org/10.1103/physrevresearch.2.032068}{Phys. Rev. Research \textbf{2},
032068(R) (2020)}.

\bibitem{ref33} L. Liu, C. Miao, H. Tang, Y.-T. Zhang, and Z. Qiao, Magnetically
controlled topological braiding with Majorana corner states in second-order topological
superconductors,
\href{https://doi.org/10.1103/physrevb.109.115413}{Phys. Rev. B \textbf{109}, 115413
(2024)}.

\bibitem{ref34} S.-R. He, X.-L. Bi, J. Fan, Z.-H. Wang, L. Li, and D.-H. Xu, Manipulating
and braiding Majorana corner modes on a kagome lattice,
\href{https://doi.org/10.1103/physrevb.111.144506}{Phys. Rev. B \textbf{111}, 144506
(2025)}.

\bibitem{ref35} Y. Shi, Q. Wang, Z.-G. Fu, P. Zhang, and N. Hao, Higher-order topological
superconductivity and electrically tunable Majorana corner modes in monolayer
Mn$X$Pb$_2$--Pb ($X=\mathrm{Se},\mathrm{Te}$) heterostructure,
\href{https://doi.org/10.1103/p1w1-rvwg}{Phys. Rev. B \textbf{114}, L020506 (2026)}.

\bibitem{ref36} L. \v{S}mejkal, J. Sinova, and T. Jungwirth, Beyond conventional
ferromagnetism and antiferromagnetism: A phase with nonrelativistic spin and crystal
rotation symmetry,
\href{https://doi.org/10.1103/physrevx.12.031042}{Phys. Rev. X \textbf{12}, 031042
(2022)}.

\bibitem{ref37} L. \v{S}mejkal, J. Sinova, and T. Jungwirth, Emerging research landscape
of altermagnetism,
\href{https://doi.org/10.1103/physrevx.12.040501}{Phys. Rev. X \textbf{12}, 040501
(2022)}.

\bibitem{ref38} C. Song, H. Bai, Z. Zhou, L. Han, H. Reichlova, J. H. Dil, J. Liu, X.
Chen, and F. Pan, Altermagnets as a new class of functional materials,
\href{https://doi.org/10.1038/s41578-025-00779-1}{Nat. Rev. Mater. \textbf{10}, 473
(2025)}.

\bibitem{ref39} S. A. A. Ghorashi, T. L. Hughes, and J. Cano, Altermagnetic routes to
Majorana modes in zero net magnetization,
\href{https://doi.org/10.1103/physrevlett.133.106601}{Phys. Rev. Lett. \textbf{133},
106601 (2024)}.

\bibitem{ref40} Y.-X. Li and C.-C. Liu, Majorana corner modes and tunable patterns in an
altermagnet heterostructure,
\href{https://doi.org/10.1103/physrevb.108.205410}{Phys. Rev. B \textbf{108}, 205410
(2023)}.

\bibitem{ref41} Y.-X. Li, Realizing tunable higher-order topological superconductors with
altermagnets,
\href{https://doi.org/10.1103/physrevb.109.224502}{Phys. Rev. B \textbf{109}, 224502
(2024)}.

\bibitem{ref42} A. Hadjipaschalis, S. A. A. Ghorashi, and J. Cano, Majoranas with a twist:
Tunable Majorana zero modes in altermagnetic heterostructures,
\href{https://doi.org/10.1103/p79l-rty6}{Phys. Rev. B \textbf{112}, 214430 (2025)}.

\bibitem{ref43} G. Z. X. Yang, Z.-T. Sun, Y.-M. Xie, and K. T. Law, Topological
altermagnetic Josephson junctions,
\href{https://doi.org/10.1038/s41535-026-00874-8}{npj Quantum Mater. \textbf{11}, 45
(2026)}.

\bibitem{ref44} B. Fu, C.-A. Li, and B. Trauzettel, Altermagnetism-induced Bogoliubov
Fermi surfaces form topological superconductivity,
\href{https://doi.org/10.1103/hz7z-m9tn}{Phys. Rev. Lett. \textbf{137}, 096604 (2026)}.

\bibitem{ref45} T. Hodge, E. Mascot, and S. Rachel, Altermagnet-superconductor
heterostructure: A scalable platform for braiding of Majorana modes,
\href{https://doi.org/10.1103/bfwd-hcm5}{APS Open Sci. \textbf{1}, L000045 (2026)}.

\bibitem{ref46} Z. Li, Z. Li, and Z. Qiao, Altermagnetism-induced topological phase
transitions in the Kane--Mele model,
\href{https://doi.org/10.1103/physrevb.111.155303}{Phys. Rev. B \textbf{111}, 155303
(2025)}.

\bibitem{ref47} G. McArdle, B. Kiraly, P. Wadley, and A. Gammon-Smith, Topological
superconductivity in altermagnetic heterostructures on a honeycomb lattice,
\href{https://arxiv.org/abs/2601.05662}{arXiv:\allowbreak\mbox{2601.05662}} [cond-mat.mes-hall] (2026).

\bibitem{ref48} C. L. Kane and E. J. Mele, Quantum spin Hall effect in graphene,
\href{https://doi.org/10.1103/physrevlett.95.226801}{Phys. Rev. Lett. \textbf{95}, 226801
(2005)}.

\bibitem{ref49} D. Zhu, B.-X. Li, and Z. Yan,
Sublattice-sensitive Majorana modes,
\href{https://doi.org/10.1103/physrevb.106.245418}{Phys. Rev. B \textbf{106}, 245418 (2022)}.

\bibitem{ref50} R. Jackiw and C. Rebbi, Solitons with fermion number $1/2$,
\href{https://doi.org/10.1103/physrevd.13.3398}{Phys. Rev. D \textbf{13}, 3398 (1976)}.

\bibitem{ref51} D. A. Ivanov,
Non-Abelian statistics of half-quantum vortices in $p$-wave superconductors,
\href{https://doi.org/10.1103/PhysRevLett.86.268}
{Phys. Rev. Lett. \textbf{86}, 268 (2001)}.

\bibitem{ref52} F. Harper, A. Pushp, and R. Roy,
Majorana braiding in realistic nanowire Y-junctions and tuning forks,
\href{https://doi.org/10.1103/PhysRevResearch.1.033207}
{Phys. Rev. Research \textbf{1}, 033207 (2019)}.

\bibitem{ref53} S. Chourasia, A. Svetogorov, A. Kamra, and W. Belzig,
Thermodynamic properties of a superconductor interfaced with an altermagnet,
\href{https://doi.org/10.1103/physrevb.111.224503}{Phys. Rev. B \textbf{111}, 224503 (2025)}.

\bibitem{ref54} S.-B. Zhang, L.-H. Hu, and T. Neupert,
Finite-momentum Cooper pairing in proximitized altermagnets,
\href{https://doi.org/10.1038/s41467-024-45951-3}{Nat. Commun. \textbf{15}, 1801 (2024)}.

\bibitem{ref55} F. Reis, G. Li, L. Dudy, M. Bauernfeind, S. Glass, W. Hanke, R. Thomale,
J. Sch\"afer, and R. Claessen, Bismuthene on a SiC substrate: A candidate for a
high-temperature quantum spin Hall material,
\href{https://doi.org/10.1126/science.aai8142}{Science \textbf{357}, 287 (2017)}.

\bibitem{ref56} F. L\"upke, D. Waters, S. C. de la Barrera, M. Widom, D. G. Mandrus, J.
Yan, R. M. Feenstra, and B. M. Hunt, Proximity-induced superconducting gap in the quantum
spin Hall edge state of monolayer WTe$_2$,
\href{https://doi.org/10.1038/s41567-020-0816-x}{Nat. Phys. \textbf{16}, 526 (2020)}.

\bibitem{ref57} J. Krempask\'y \emph{et~al.}, Altermagnetic lifting of Kramers spin
degeneracy,
\href{https://doi.org/10.1038/s41586-023-06907-7}{Nature \textbf{626}, 517 (2024)}.

\bibitem{ref58} S. Lee, S. Lee, S. Jung, J. Jung, D. Kim, Y. Lee, B. Seok, J. Kim, B. G.
Park, L. \v{S}mejkal, C.-J. Kang, and C. Kim, Broken Kramers degeneracy in altermagnetic
MnTe,
\href{https://doi.org/10.1103/physrevlett.132.036702}{Phys. Rev. Lett. \textbf{132},
036702 (2024)}.

\bibitem{ref59} S. Reimers, L. Odenbreit, L. \v{S}mejkal, V. N. Strocov, P. Constantinou,
A. B. Hellenes, R. Jaeschke Ubiergo, W. H. Campos, V. K. Bharadwaj, A. Chakraborty, T.
Denneulin, W. Shi, R. E. Dunin-Borkowski, S. Das, M. Kl\"aui, J. Sinova, and M. Jourdan,
Direct observation of altermagnetic band splitting in CrSb thin films,
\href{https://doi.org/10.1038/s41467-024-46476-5}{Nat. Commun. \textbf{15}, 2116 (2024)}.

\bibitem{ref60} B. Jiang \emph{et~al.}, A metallic room-temperature $d$-wave altermagnet,
\href{https://doi.org/10.1038/s41567-025-02822-y}{Nat. Phys. \textbf{21}, 754 (2025)}.

\bibitem{ref61} O. J. Amin \emph{et~al.}, Nanoscale imaging and control of altermagnetism
in MnTe,
\href{https://doi.org/10.1038/s41586-024-08234-x}{Nature \textbf{636}, 348 (2024)}.

\bibitem{ref62} H. Ren, F. Pientka, S. Hart, A. T. Pierce, M. Kosowsky,
L. Lunczer, R. Schlereth, B. Scharf, E. M. Hankiewicz, L. W. Molenkamp,
B. I. Halperin, and A. Yacoby, Topological superconductivity in a phase-controlled
Josephson junction,
\href{https://doi.org/10.1038/s41586-019-1148-9}{Nature \textbf{569}, 93 (2019)}.

\bibitem{ref63} K. Yoshizawa, R. Okugawa, and T. Tohyama,
Topological superconducting phases in a strained altermagnet-superconductor
heterostructure,
\href{https://arxiv.org/abs/2609.30791}{arXiv:\allowbreak\mbox{2609.30791}}
[cond-mat.supr-con] (2026).

\bibitem{ref64} K. S. Novoselov, A. K. Geim, S. V. Morozov, D. Jiang,
Y. Zhang, S. V. Dubonos, I. V. Grigorieva, and A. A. Firsov, Electric field effect
in atomically thin carbon films,
\href{https://doi.org/10.1126/science.1102896}{Science \textbf{306}, 666 (2004)}.

\bibitem{ref65} M. Ezawa, A topological insulator and helical zero mode in
silicene under an inhomogeneous electric field,
\href{https://doi.org/10.1088/1367-2630/14/3/033003}{New J. Phys. \textbf{14}, 033003 (2012)}.

\bibitem{ref66} S. Tewari and J. D. Sau, Topological invariants for spin-orbit coupled
superconductor nanowires,
\href{https://doi.org/10.1103/physrevlett.109.150408}{Phys. Rev. Lett. \textbf{109},
150408 (2012)}.

\bibitem{ref67} A. P. Schnyder, S. Ryu, A. Furusaki, and A. W. W. Ludwig, Classification
of topological insulators and superconductors in three spatial dimensions,
\href{https://doi.org/10.1103/physrevb.78.195125}{Phys. Rev. B \textbf{78}, 195125
(2008)}.

\bibitem{ref68} P. Boross and A. P\'alyi, Braiding-based quantum control of a Majorana
qubit built from quantum dots,
\href{https://doi.org/10.1103/physrevb.109.125410}{Phys. Rev. B \textbf{109}, 125410
(2024)}.

\bibitem{ref69} M. Nitsch, V. Svensson, W. Samuelson, K. Nestmann, J. Danon,
K. Flensberg, R. Seoane Souto, and M. Leijnse,
Adiabatic non-Abelian braiding of imperfect Majorana bound states,
\href{https://doi.org/10.1103/3vcb-pxp6}{Phys. Rev. B \textbf{113}, L201408 (2026)}.

\bibitem{ref70} C. Knapp, M. Zaletel, D. E. Liu, M. Cheng, P. Bonderson, and C. Nayak,
The nature and correction of diabatic errors in anyon braiding,
\href{https://doi.org/10.1103/physrevx.6.041003}{Phys. Rev. X \textbf{6}, 041003 (2016)}.

\end{thebibliography}
\end{document}